\documentclass[fleqn,usenatbib]{mnras}

\usepackage{newtxtext,newtxmath}

\usepackage[T1]{fontenc}

\DeclareRobustCommand{\VAN}[3]{#2}
\let\VANthebibliography\thebibliography
\def\thebibliography{\DeclareRobustCommand{\VAN}[3]{##3}\VANthebibliography}

\usepackage{graphicx}	
\usepackage{amsmath}	
\usepackage{hyperref}
\usepackage{multirow}
\usepackage{papermacs}

\newcommand{\nnni}{\ensuremath{\nuc{56}{Ni}}\xspace}

\newcommand\arcdeg{\mbox{$^\circ$}}%

\makeatletter

\newcommand\Autoref[1]{\@first@ref#1,@}
\def\@throw@dot#1.#2@{#1}
\def\@set@refname#1{
    \edef\@tmp{\getrefbykeydefault{#1}{anchor}{}}%
    \def\@refname{\@nameuse{\expandafter\@throw@dot\@tmp.@autorefname}s}%
}
\def\@first@ref#1,#2{%
  \ifx#2@\autoref{#1}\let\@nextref\@gobble
  \else%
    \@set@refname{#1}
    \@refname~\ref{#1}
    \let\@nextref\@next@ref
  \fi%
  \@nextref#2%
}
\def\@next@ref#1,#2{%
   \ifx#2@ and~\ref{#1}\let\@nextref\@gobble
   \else, \ref{#1}
   \fi%
   \@nextref#2%
}

\makeatother

\graphicspath{{./figs/}}

\title[SN~2021fxy: MUV Suppression in SNe Ia]{SN~2021fxy: Mid-Ultraviolet Flux Suppression is a Common Feature of Type Ia Supernovae}

\author[J.~M.~DerKacy et al.]{
J.~M.~DerKacy,$^{1,2}$\thanks{E-mail: jmderkacy@vt.edu}
S.~Paugh,$^{1}$
E.~Baron,$^{1,3}$
P.~J.~Brown,$^{4}$
C.~Ashall,$^{2}$
C.~R.~Burns,$^{5}$
E.~Y.~Hsiao,$^{6}$
\newauthor
S.~Kumar,$^{6}$
J.~Lu,$^{6}$
N.~Morrell,$^{7}$
M.~M.~Phillips,$^{7}$
M.~Shahbandeh,$^{6,}$
B.~J.~Shappee,$^{9}$
M.~D.~Stritzinger,$^{10}$
\newauthor
M.~A.~Tucker,$^{11,12,13}$
Z.~Yarbrough,$^{1,14}$
K.~Boutsia,$^{7}$
P.~Hoeflich,$^{6}$
L.~Wang,$^{4}$
L.~Galbany,$^{15,16}$
\newauthor
E.~Karamehmetoglu,$^{10}$
K.~Krisciunas,$^{4}$
P.~Mazzali,$^{17}$
A.~L.~Piro,$^{5}$
N.~B.~Suntzeff,$^{4}$
A.~Fiore,$^{18,19,20}$
\newauthor
C.~P.~Guti\'errez,$^{21,22}$
P.~Lundqvist,$^{23}$
and A.~Reguitti$^{24,25,20}$
\\
$^{1}$Homer L. Dodge Department of Physics and Astronomy, University of Oklahoma, 440 W. Brooks, Norman, OK 73019-2061, USA \\
$^{2}$Department of Physics, Virginia Tech, 850 West Campus Drive, Blacksburg VA, 24061, USA \\
$^{3}$Hamburger Sternwarte, Gojenbergsweg 112, 21029 Hamburg, Germany \\
$^{4}$George P. and Cynthia Woods Mitchell Institute for Fundamental Physics and Astronomy, Department of Physics and Astronomy, Texas A\&M University, \\
College Station, TX 77843, USA \\
$^{5}$The Observatories of the Carnegie Institution for Science, 813 Santa Barbara Street, Pasadena, CA 91101, USA \\
$^{6}$Department of Physics, Florida State University, 77 Chieftan Way, Tallahassee, FL 32306, USA \\
$^{7}$Carnegie Observatories, Las Campanas Observatory, Casilla 601, La Serena, Chile \\
$^{8}$Space Telescope Science Institute, 3700 San Martin Drive, Baltimore, MD 21218-2410, USA \\
$^{9}$Institute for Astronomy, University of Hawaii, 2680 Woodlawn Drive, Honolulu, HI 96822, USA \\
$^{10}$Department of Physics and Astronomy, Aarhus University, Ny Munkegade 120,  DK-8000 Aarhus C, Denmark \\
$^{11}$CCAPP Fellow, Center for Cosmology and Astroparticle Physics, The Ohio State University, 191 West Woodruff Ave, Columbus, OH, USA \\
$^{12}$Department of Astronomy, The Ohio State University, 140 West 18th Avenue, Columbus, OH, USA \\
$^{13}$Department of Physics, The Ohio State University, 191 West Woodruff Ave, Columbus, OH, USA \\
$^{14}$Department of Physics \& Astronomy, Louisiana State University, Baton Rouge, LA, USA \\
$^{15}$Institute of Space Sciences (ICE, CSIC), Campus UAB, Carrer de Can Magrans, s/n, E-08193 Barcelona, Spain \\
$^{16}$Institut d’Estudis Espacials de Catalunya (IEEC), E-08034 Barcelona, Spain \\
$^{17}$Astrophysics Research Institute, Liverpool John Moores University, IC2, Liverpool Science Park, 146  Brownlow Hill, Liverpool L3 5RF, UK \\
$^{18}$European Centre for Theoretical Studies in Nuclear Physics and Related Areas (ECT*), Fondazione Bruno Kessler, Trento, Italy \\
$^{19}$INFN-TIFPA, Trento Institute for Fundamental Physics and Applications, Via Sommarive 14, I-38123 Trento, Italy \\
$^{20}$INAF – Osservatorio Astronomico di Padova, Vicolo dell'Osservatorio 5, 35122 Padova, Italy \\
$^{21}$Finnish Centre for Astronomy with ESO (FINCA), FI-20014 University of Turku, Finland \\
$^{22}$Tuorla Observatory, Department of Physics and Astronomy, FI-20014 University of Turku, Finland \\
$^{23}$Oskar Klein Centre, Department of Astronomy, Stockholm University, Albanova University Centre, SE-106 91 Stockholm, Sweden \\
$^{24}$Instituto de Astrof\`{i}sica – Universidad Andres Bello, Avda. Rep\'{u}blica 252, 8320000, Santiago, Chile \\
$^{25}$Millennium Institute of Astrophysics, Nuncio Monsenor S\'{o}tero Sanz 100, Providencia, 8320000, Santiago, Chile \\
}

\date{Accepted XXX. Received YYY; in original form ZZZ}

\pubyear{2022}

\begin{document}
\label{firstpage}
\pagerange{\pageref{firstpage}--\pageref{lastpage}}
\maketitle

\begin{abstract}
We present ultraviolet (UV) to near-infrared (NIR) observations and analysis of 
the nearby Type Ia supernova SN~2021fxy. Our observations include UV photometry 
from {\it Swift}/UVOT, UV spectroscopy from {\it HST}/STIS, and high-cadence 
optical photometry with the Swope 1-m telescope capturing intra-night rises 
during the early light curve. Early $B-V$ colours show SN~2021fxy is the first 
``shallow-silicon" (SS) SN~Ia to follow a red-to-blue evolution, compared to 
other SS objects which show blue colours from the earliest observations.
Comparisons to other spectroscopically normal SNe~Ia with {\it HST} UV spectra 
reveal SN~2021fxy is one of several SNe~Ia with flux suppression in the mid-UV. 
These SNe also show blue-shifted mid-UV spectral features and strong high-velocity 
\ion{Ca}{ii} features. One possible origin of this mid-UV suppression is the 
increased effective opacity in the UV due to increased line blanketing from 
high velocity material, but differences in the explosion mechanism cannot be 
ruled out. Among SNe~Ia with mid-UV suppression, SNe~2021fxy and 2017erp show 
substantial similarities in their optical properties despite belonging to 
different Branch subgroups, and UV flux differences of the same order as those 
found between SNe~2011fe and 2011by. Differential comparisons to multiple sets 
of synthetic SN~Ia UV spectra reveal this UV flux difference likely originates 
from a luminosity difference between SNe~2021fxy and 2017erp, and not differing 
progenitor metallicities as suggested for SNe~2011by and 2011fe. These 
comparisons illustrate the complicated nature of UV spectral formation, and the 
need for more UV spectra to determine the physical source of SNe~Ia UV diversity.
\end{abstract}

\begin{keywords}
supernovae: general -- supernovae: individual (SN~2021fxy, SN~2013dy, SN~2017erp, ASASSN-14lp)
\end{keywords}



\section{Introduction} \label{sec:intro}

Type Ia supernovae (SNe~Ia) are important astrophysical objects because of 
their utility as ``standardizable candles'' for cosmological studies. The 
empirical Phillips relation \citep{Phillips1993,Phillips:1999} allows them 
to serve as cosmological probes, revealing the accelerating expansion rate 
of the universe \citep{Riess1998,Perlmutter1999}. 

While it is well established that SNe~Ia are explosions of a primary 
carbon and oxygen (C/O) white dwarves in binary systems \citep{Hoyle1960}, 
the full nature of the progenitor system, including the identity of the 
secondary star and the explosion mechanism, are still unclear 
\citep[for a review see][]{Maoz:2014}. 
In the single-degenerate (SD) scenario, the companion is either a 
main-sequence star or an evolved, non-degenerate companion like a red 
giant or He-star \citep{Whelan1973}. In the double-degenerate (DD) 
scenario, the companion is also a white dwarf, where the explosion is 
triggered by the merger or interaction of the two WDs 
\citep{Iben1984,Webbink1984}. More recent work has suggested that some 
SNe~Ia may potentially originate from the merger of a WD with the core 
of an evolved star; known as the core-degenerate scenario 
\citep{Kashi2011,Soker2014}. A wide range of explosion mechanisms have 
been proposed, including pure deflagrations \citep{Nomoto1984}, 
detonation-to-deflagration transitions (DDTs; also referred to as delayed 
detonations) \citep{Hoeflich1995,Hoeflich2002,Hoeflich2006}, surface 
helium detonations 
\citep{Thielemann1986,Woosley:1994:Weaver:HeDet,Livne:1995:Arnett:HeDet,Shen2018,Polin2019}, 
and detonations triggered by the ``violent" mergers of two WDs before 
they are able to fully merge \citep{Rosswog2009,Pakmor2010,Pakmor2012}.
There is no consensus as to whether SNe~Ia occur by any or all of
these proposed mechanisms, making the SNe~Ia progenitor/explosion
mechanism problem extremely complex.

Several different schemes have been developed to characterize the 
observed diversity of SNe~Ia near maximum light. \citet{Branch2006} 
subdivide SNe~Ia into four groups: core-normal~(CN), shallow silicon~(SS), 
broad line~(BL), and cool~(CL); based upon the pseudo-equivalent 
widths~(pEWs) of the \ion{Si}{ii} $\lambda 5972$ and $\lambda 6355$ lines 
in their spectra near maximum light. Recent work with larger samples have 
shown these groups to be statistically robust \citep{Burrow2020} and are 
potentially related to differences in the progenitor systems and/or 
explosion mechanisms \citep{Polin2019}. \citet{Wang2009b} divide SNe~Ia 
into ``Normal'' and ``High Velocity (HV)'' groups based on the velocity 
of the \ion{Si}{ii} $\lambda6355$ absorption minimum near $B$-band maximum 
light, with the HV objects showing redder $B-V$ colours and less scatter in 
peak luminosity and luminosity decline rate ($\Delta m_{15}$) relative to 
Normal SNe~Ia. \citet{Benetti2005} divide the SNe~Ia population into 
``Faint'', ``Low Velocity Gradient'' and ``High Velocity Gradient'' 
groupings based on the combination of their decline rate in $B$-band and 
the rate of change in their \ion{Si}{ii} velocities. 

Early observations are key to determining the connections between the
observed SNe~Ia diversity and different progenitor scenarios and
explosion mechanisms.  Early observations can probe the physical
properties of the system, including constraints on the size of the WD
progenitor \citep{Piro2010,Nugent2011,Bloom2012}, the properties of 
secondary star \citep{Kasen2010,Maeda2014}, and the distribution of 
any circumstellar material~(CSM) \citep{Lundqvist2020}. Early photometric 
observations of nearby SNe~Ia discovered within hours of explosion  
reveal that some SNe~Ia (such as SNe~2017cbv, 2018oh/ASASSN-18bt, 2019np, 
and 2021aefx) show an early excess or ``blue bump'' at early times 
\citep{Hosseinzadeh2017,Li2019,Sai2022,Ashall2022,Hosseinzadeh2022}.  
However, there are multiple potential origins of these early time excesses, 
including outward mixing of $^{56}$Ni in the ejecta 
\citep{Piro:2016,Shappee2019}, production of radioactive material 
in the detonation of the helium shell \citep{Dimitriadis2019,Polin2019}, 
interaction with the companion \citep{Kasen2010,Maeda2014,Dimitriadis2019}, 
and rapid velocity evolution of broad, high-velocity spectroscopic features 
\citep{Ashall2022}. 
Attempts to probe the companion interaction in the radio have yet to 
detect a companion interaction, but have yielded information on the
nearby circumstellar environment in the first days after explosion and
provided constraints on the wind properties of the progenitor 
\citep{Lundqvist2020,Hosseinzadeh2022}.

Colour curves derived from photometry of SNe~Ia obtained within 
the first few days after explosion reveal at least two different 
populations, distinguishable by their $B-V$ behavior
\citep{Stritzinger2018}. Of the two most populous groups, the first 
exhibits blue colours from the earliest epochs, while the other group 
starts out red and rapidly becomes bluer, with both groups showing 
indistinguishable colours roughly 6 days after first light. These groups 
are not replicated in $g-r$ colours \citep{Bulla2020}. Spectroscopic 
observations within these first few days after explosion probe the 
outermost ejecta layers where differences between models of SNe~Ia are 
largest. For example, DDT models show unburned carbon, oxygen, and possibly 
silicon in the outermost layers \citep{Hoeflich2017} compared to the large 
amounts of $^{44}$Ti and $^{56}$Ni which are the expected by-products of 
models relying upon surface helium detonations \citep{Jiang2017}. In the
models of \cite{Polin2019} these by-products produce significant line
blanketing, resulting in red colours at early times. Early spectra often 
show high-velocity features (HVFs), which typically disappear before
maximum light but may be ubiquitous among SNe Ia \citep{Mazzali2005}.

SNe~Ia diversity increases as one moves from optical to ultraviolet~(UV) 
wavelengths. Photometrically, SNe~Ia can be divided into
two groups based on their near-UV~(NUV) colours; the NUV-blue group, whose
members have low velocity gradients of their \ion{Si}{ii} $\lambda6355$ lines
and conspicuous \ion{C}{ii} lines, and the NUV-red class, whose members have
more diverse \ion{Si}{ii} velocity gradients and typically lack the \ion{C}{ii}
lines \citep{Milne2013}. The fraction of events belonging to each
group varies by redshift, making it difficult to incorporate UV
data of SNe~Ia into cosmological analyses \citep{Milne2015,Brown2017}.
Spectroscopically, SNe~Ia can show drastic differences in the UV
despite being almost identical in the optical and near-infrared~(NIR). 
The best example of this are the ``twin'' SNe~2011by and 2011fe 
\citep{Foley2013}.  

Theoretical efforts to better understand UV
spectral formation have focused on the impacts of three key variables:
(1) metallicity -- increases in the progenitor metallicity
strengthen line blanketing in the UV and result in lower fluxes
\citep{Lentz2000}, (2) density structure -- shallower density
profiles produce UV spectra with lower flux values and fewer features
\citep{Sauer2008,Hachinger2013,Mazzali2014}, and (3) luminosity --
which induces temperature variations that change both the shape of the
underlying continuum and the strength, shape, and location of spectral
features \citep{Walker2012,DerKacy2020}.

In this work, we present observations and analysis of SN~2021fxy, an
NUV-red SN~Ia discovered roughly 2 days after explosion and for which 
we obtained multiple \textit{HST}/STIS UV spectra in addition to a 
comprehensive multi-band follow-up effort by the Precision
Observations of Infant Supernova Explosions 
(\href{https://poise.obs.carnegiescience.edu/}{POISE}, \citealp{Burns2021}) 
collaboration. \autoref{sec:obs} details these photometric and
spectroscopic follow-up observations, followed by a detailed analysis in
\autoref{sec:analysis}, including comparisons to the sample of spectroscopically 
normal SNe~Ia with {\it HST} UV spectra near maximum light. In 
\autoref{sec:discussion} we compare SNe~2021fxy and 2017erp, both of which 
are well-observed NUV-red SNe~Ia with {\it HST}/STIS spectra, and who show 
nearly identical optical properties. We discuss the potential causes of 
their observed differences in the context of the UV diversity of SNe~Ia 
and what they reveal about the origins of this diversity. We summarize our 
conclusions in \autoref{sec:conclusion}.

\section{Observations} \label{sec:obs}

\subsection{Discovery}

\begin{figure}
  \centering
  \includegraphics[trim=0cm 0cm 1.5cm 0.5cm,clip=True,width=\columnwidth]{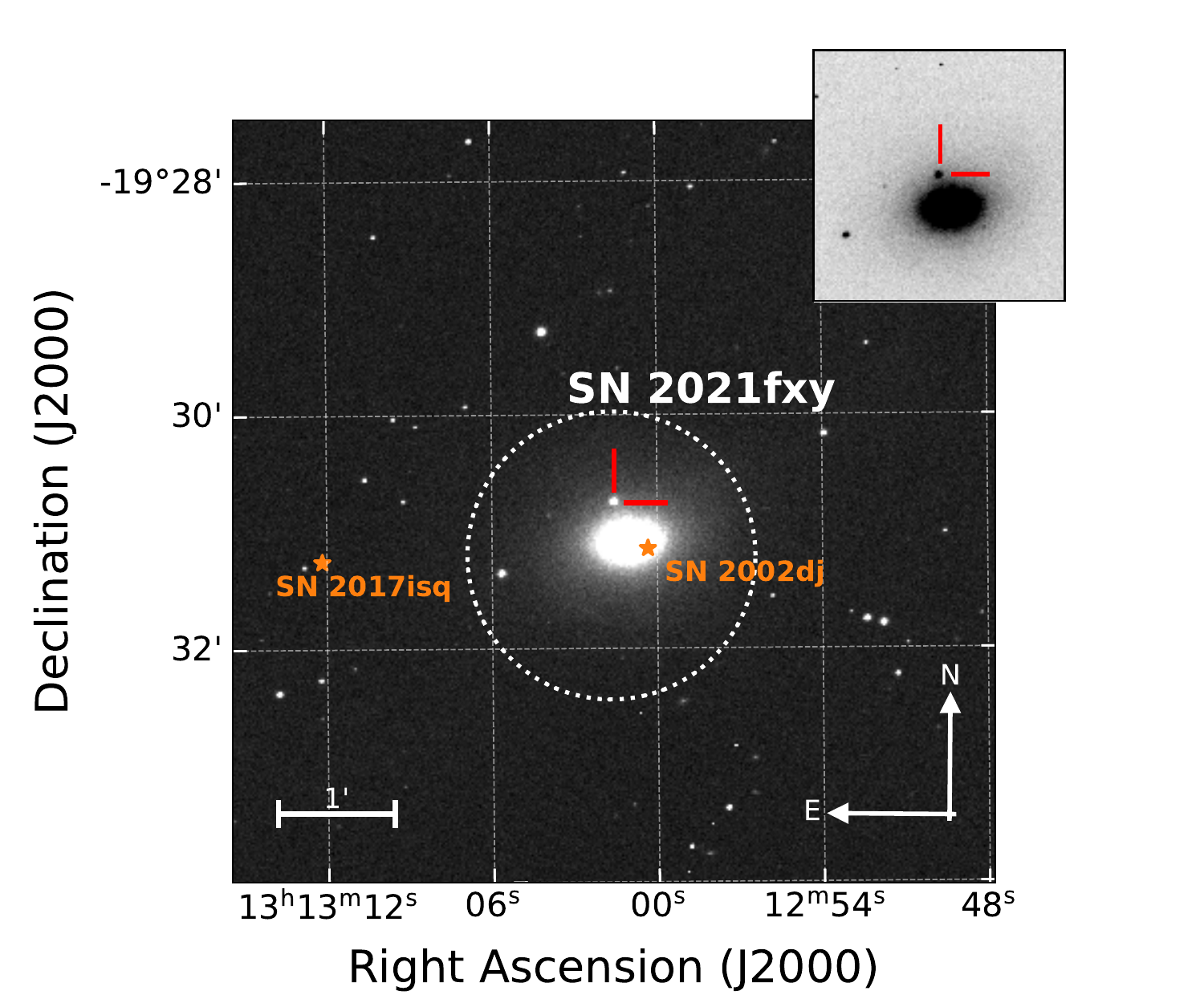}
  \caption{An $r$-band image of NGC~5018 taken the night of 2021-03-24 with 
  the Swope~1-m telescope at Las Campanas Observatory, Chile. SN~2021fxy is 
  highlighted in both the main panel and the false-colour inset with the red 
  cross-hairs. The orange stars indicate the locations of sibling SN~2002dj
  and potential sibling SN~2017isq.} 
  \label{fig:finder}
\end{figure}

SN~2021fxy was discovered on 2021 March 17.75 by Koichi Itagaki at 
$m = 16.9$ mag in a clear filter \citep{Itagaki2021}, and classified as 
a young SN~Ia the following night \citep{Jha2021}. SN~2021fxy is located 
at $\alpha = 13^{h}13^{m}01^{s}.570$, $\delta=-19\arcdeg30\arcmin45\arcsec.18$, 
which is $19\arcsec.8$ North and $8\arcsec.1$ East from the centre of the host 
galaxy, NGC~5018 (see \autoref{fig:finder}). The most constraining last 
non-detection comes from ASAS-SN on 2021 March 15.45 in the $g$-band at 
$g > 18.01$ mag, which was retrieved from the ASAS-SN Sky Patrol 
Database\footnote{\url{https://asas-sn.osu.edu/}} 
\citep{Shappee2014,Kochanek2017}. This implies SN~2021fxy was discovered 
no later than 2.27 days after explosion, assuming there is no dark phase.

NGC 5018 is classified as an E3 according to \citet{deVaucouleurs1991}, at a 
redshift $z = 0.0094$ \citep{Rothberg2006}. Correcting the velocities for 
local motions based on \citet{Mould2000} and assuming a $H_0=73$ km s$^{-1}$ 
Mpc$^{-1}$ \citep{Riess2022}, this redshift results in a Hubble flow distance of 
$38.4\pm2.7$ Mpc, corresponding to a distance modulus of  (m-M) $= 32.92 \pm 0.15$ 
mag. SN~2021fxy is at least the second SNe~Ia discovered in NGC 5018, along with
the well-studied 02bo-like SN~2002dj \citep{Pignata2008}. Another potential SN~Ia
sibling, SN~2017isq, was discovered roughly one month after maximum light at 
an estimated separation of 30 kpc from NGC~5018; its closest potential host 
\citep{Tonry2017,Benetti2017}. Important information on SN~2021fxy and NGC~5018
can be found in \autoref{tab:props}.

\begin{table}
\centering
\caption{Properties of SN~2021fxy and NGC 5018 \label{tab:props}}
\begin{tabular}{ccc}
  \hline
  Parameter & Value & Source \\
  \hline
  SN~2021fxy: & & \\
  R.A. [J2000] & $13^{h}13^{m}01^{s}.570$ & (1) \\
  Dec. [J2000] & $-19\arcdeg30\arcmin45\arcsec.18$ & (1) \\
  $t(B_{max})$ [MJD] & $59305.12 \pm 0.34$ & This Work \\
  $B_{max}$ [mag]$^{a}$ & $13.56 \pm 0.07$ & This Work \\
  $E(B-V)_{MW}$ [mag] & $0.084$ & (2) \\ 
  $E(B-V)_{Host}$ [mag] & $0.02 \pm 0.06$ & This Work \\ 
  $s_{BV}$ & $ 0.99 \pm 0.03$ & This Work \\
  $\Delta m_{15}(B)$ [mag]$^{b}$ & $1.05 \pm 0.06$ & This Work \\
  \hline
  NGC 5018: & & \\
  R.A. [J2000] & $13^{h}13^{m}01^{s}.70$ & (3) \\
  Dec. [J2000] & $-19\arcdeg31\arcmin12\arcsec.8$ & (3) \\
  Morphology & E3 & (4) \\
  Heliocentric Velocity, & \multirow{2}{*}{$2816 \pm 1$} & \multirow{2}{*}{(3)}\\
  $v$ [\kmps]&  &  \\
  $z$ & $0.0094$ & (3) \\
  $\mu$ & $32.92 \pm 0.15$ & (3) \\
  \hline
\end{tabular} \\
$^{a}$~Apparent magnitude, corrected for extinction.
$^{b}$~Value from \texttt{SNooPy} fit with \texttt{EBV\_model2} to SN~2021fxy light curve. \\
References -- (1) \citet{Itagaki2021}, (2) \citet{Schlafly2011}, (3) \citet{Rothberg2006}, 
(4) \citet{deVaucouleurs1991}.
\end{table}

\subsection{Photometric Follow-up}

\subsubsection{Ground-based Photometry}

The POISE collaboration began a multi-band ($uBVgri$) follow-up campaign 
of SN~2021fxy using the Swope 1-m telescope at Las Campanas Observatory at 
2021 March 18.11, just 0.39 days after discovery, placing our first 
observations no later than 2.66 days after last non-detection. During the 
early rise, two sets of observations were taken per night to capture rapid, 
intra-night evolution. Roughly one week after discovery, the cadence was 
reduced to one set of observations per night. Observations from the Swope 
were reduced and analysed according to the procedures of 
\cite{Krisciunas2017} and \cite{Phillips2019}. 

Some post-maximum photometric observations in $BVgri$ bands were taken 
using the Las Cumbres Observatory global 1-m telescope network (LCOGT) 
as part of the Aarhus-Barcelona FLOWS project.\footnote{\url{https://flows.phys.au.dk/}} 
This data was reduced with the BANZAI pipeline \citep{McCully2018} and 
calibrated using the local sequence photometry from the Swope observations, 
assuming zero colour terms. 

Photometry obtained from both telescopes are presented in the CSP-II 
natural system in \autoref{sec:phot}. For data obtained with the 
Swope 1-m, S-corrections \citep{Stritzinger2002} between the CSP-II 
and CSP-I systems are quite small ($\lesssim 0.01$ mag) in all but 
the $u$-band, and are smaller than the typical photometric uncertainty
in all bands. Therefore, we conclude that the CSP-I and CSP-II systems 
are effectively the same (only the CCD was changed between the two 
projects; \citealp[][Suntzeff et al., in preparation]{Phillips2019}).
Similarly, because the data obtained with the LCOGT network were 
calibrated using local sequence photometry from the Swope images, its 
natural system will be only marginally different than CSP-II; and we
therefore treat the data as it if is on the CSP-II system. Unaccounted 
for errors between the LCOGT natural system and the CSP-II system
arising from slightly different photometric transmission functions and 
differences between individual telescopes within the LCOGT network are 
expected to be small ($\lesssim 3\%$).

\subsubsection{\textit{Swift} Photometry}

{\it Swift} observations were first triggered as part of the {\it Swift} GI program 
``Maximizing Swift's Impact With The Global Supernova Project'' (PI: Howell).  
Observations began on 2021 March 18.25 (2.80 days after last non-detection). 
Due to the brightness of the underlying host galaxy, some optical observations 
were made in a hardware mode with a faster readout to reduce the effect of 
coincidence loss. Photometry was computed with the Swift Optical 
Ultraviolet Supernova Archive \citep[SOUSA;][]{Brown_2014} pipeline using 
the 2020 update to the time-dependent sensitivity and aperture corrections 
calculated in 2021. No subtraction of the host-galaxy flux has been 
performed due to the lack of pre-explosion images.

\subsection{Spectroscopic Follow-up}

\begin{table}
\centering
\caption{Log of spectroscopic observations \label{tab:spec_log}} 
\begin{tabular}{@{}c@{\:}c@{\:}c@{\:}c@{\:}c@{}}
  \hline
  Date (UT) & MJD & Epoch$^{a}$ & Obs. Range (\AA) & Telescope/Instrument \\
  \hline
  \multicolumn{5}{c}{Ultraviolet Spectra} \\
  \hline
  2021 Mar 29.8 & 59302.83 & $-2.3$ & 1600-5600 & \textit{HST}/STIS \\
  2021 Apr 01.6 & 59305.55 & $+0.4$ & 1600-5600 & \textit{HST}/STIS \\
  2021 Apr 08.4 & 59312.43 & $+7.2$ & 1600-5600 & \textit{HST}/STIS \\
  \hline
  \multicolumn{5}{c}{Optical Spectra} \\
  \hline
  2021 Mar 18.1 & 59291.11 & $-13.9$ & 3496-9372 & SALT/RSS$^{b}$ \\
  2021 Mar 25.1 & 59299.09 & $-5.9$ & 3398-9674 & NOT/ALFOSC \\
  2021 Apr 02.4 & 59306.38 & $+1.3$ & 3400-9840 & ARC/DIS \\
  2021 Apr 03.1 & 59307.05 & $+2.0$ & 3397-9673 & NOT/ALFOSC \\
  2021 Apr 04.3 & 59308.30 & $+3.2$ & 3400-9840 & ARC/DIS \\
  2021 Apr 06.2 & 59310.21 & $+5.1$ & 3400-9840 & ARC/DIS \\
  2021 Apr 08.5 & 59312.49 & $+7.3$ & 3787-9100 & UH88/SNIFS \\
  2021 Apr 12.1 & 59316.11 & $+10.9$ & 3400-9683 & NOT/ALFOSC \\
  2021 Apr 13.3 & 59317.31 & $+12.1$ & 3400-9864 & ARC/DIS \\
  2021 Apr 17.1 & 59321.13 & $+15.9$ & 3439-9412 & Magellan (Clay)/MIKE \\
  2021 Apr 17.2 & 59321.24 & $+16.0$ & 3816-10632 & Magellan (Clay)/LDSS3 \\
  2021 Apr 20.0 & 59324.04 & $+18.7$ & 3401-9639 & NOT/ALFOSC \\
  2021 Apr 20.1 & 59324.11 & $+18.9$ & 3707-9290 & Magellan (Clay)/LDSS3 \\
  2021 May 03.9 & 59337.93 & $+32.5$ & 3398-9653 & NOT/ALFOSC \\
  2021 May 16.9 & 59350.91 & $+45.4$ & 3689-9687 & NOT/ALFOSC \\
  \hline
  \multicolumn{5}{c}{Near-infrared Spectra} \\
  \hline
  2021 Apr 02.4 & 59306.43 & $+1.3$ & 6905-25701 & IRTF/SpeX \\
  2021 Apr 19.6 & 59323.56 & $+18.2$ & 6909-25714 & IRTF/SpeX \\
  2021 May 10.2 & 59344.24 & $+38.7$ & 6868-25384 & IRTF/SpeX \\
  \hline
\end{tabular} \\
$^{a}$~Rest frame days relative to $B$-band maximum of MJD$=59305.12$. \\
$^{b}$~Retrieved from TNS \citep{Jha2021}.
\end{table}

\subsubsection{Optical Spectroscopy}

Optical spectroscopic follow-up observations, covering $-5.9$ days 
to $+45.4$ rest frame days relative to $B$-band maximum were made 
with a global network of telescopes and instruments, including the
Dual Imaging Spectrgrpah (DIS) 
on the Astrophysical Research Consortium 3.5-meter telescope at 
Apache Point Observatory (ARC 3.5-m), ALFOSC on the Nordic Optical 
Telescope (NOT) by the NUTS2 
collaboration\footnote{\url{https://nuts.sn.ie/}}, the Supernova 
Integral Field Spectrogaph (SNIFS; \citealt{Lantz2004}) on the University 
of Hawaii 2.2-meter telescope (UH88), and both the Magellan Inamori Kyocera 
Echelle (MIKE) and the Low Dispersion Survey Spectrograph (LDSS3) 
instruments on the Landon T. Clay (Magellan) Telescope at Las Campanas 
Observatory. A classification spectrum taken with the Robert Stobie 
Spectrograph (RSS) on the South African Large Telescope (SALT) at $-13.9$ days 
was retrieved from Transient Name Server\footnote{\url{https://www.wis-tns.org/}} 
and is also included here \citep{Jha2021}.

Spectra taken with the ARC 3.5-m were reduced using standard 
IRAF\footnote{IRAF is distributed by the National Optical Astronomy Observatories, 
which are operated by the Association of Universities for Research in Astronomy, 
Inc., under cooperative agreement with the National Science Foundation (NSF).} 
methods including bias subtraction, flat fielding, cosmic ray removal using L.A. 
Cosmic\footnote{\url{http://www.astro.yale.edu/dokkum/lacosmic/}} package 
\citep{vanDokkum2001}, and flux calibration from a spectrophotometric standard 
star taken at a similar airmass that same night. \nocite{Tody1986,Tody1993}
The SNIFS spectrum is traced, extracted, and calibrated with custom Python 
routines \citep{Tucker2022} and atmospheric attenuation is corrected using 
the results of \cite{Buton2013}. The spectrum taken with Magellan/MIKE was 
processed through a combination of IRAF echelle tasks and the 
``mtools"\footnote{\url{http://www.lco.cl/?epkb_post_type_1=iraf-mtools-package}} 
package, specially developed for the reduction of MIKE spectra. A flux 
standard obtained during the same night of the observations was used as flux 
calibrator. Flux calibration was also checked with a low resolution Magellan/LDSS3 
spectrum of SN~2021fxy obtained during the same night as the MIKE observation.

\subsubsection{HST Spectroscopy}

UV spectroscopy of SN~2021fxy  with the \textit{Hubble Space Telescope} (HST) 
equipped with the Space Telescope Imaging Spectrograph (STIS) using the mid-UV G230L 
and the near-UV/optical G430L gratings was triggered by the program ``Red or 
Reddened Supernovae? Understanding the Ultraviolet Differences of Normal 
Standard Candles'' (PI: Brown; ID: 16221). Observations were scheduled for 29/30 
Mar, 01 Apr, 03 Apr, and 08 Apr. Some observations on 01 Apr and all 03 Apr were 
unusable because of a guide star acquisition failure. Reduced spectra were 
obtained from the Mikulski Archive for Space Telescopes (MAST\footnote{
\url{https://science.nasa.gov/astrophysics/astrophysics-data-centers/multimission-
archive-at-stsci-mast}}). The multiple spectra from both grisms were combined
using a weighted average within a bin of 5 \AA.

\subsubsection{NIR Spectroscopy}

The NIR spectra of SN~2021fxy were obtained with the SpeX 
\citep{2003PASP..115..362R} spectrograph installed on the 3.0-m 
NASA Infrared Telescope Facility (IRTF) on three epochs (2021-04-02, 
2021-04-19 and 2021-05-10). The spectra were taken in both the PRISM 
and SXD mode with a slit size of 0.5$\times$15\arcsec. The spectra were taken 
using the classic ABBA technique, and were reduced utilizing the 
\texttt{Spextool} software package \citep{2004PASP..116..362C}. The 
telluric absorption corrections were done using the \texttt{XTELLCOR} 
software. Complete details of the reduction procedure can be found in
\cite{Hsiao2019}. The complete log of spectroscopic observations is given 
in \autoref{tab:spec_log}.

\section{Analysis} \label{sec:analysis}

\subsection{Light Curve Analysis} \label{sec:lc_analysis}

\begin{figure*}
  \centering
  \includegraphics[width=\textwidth]{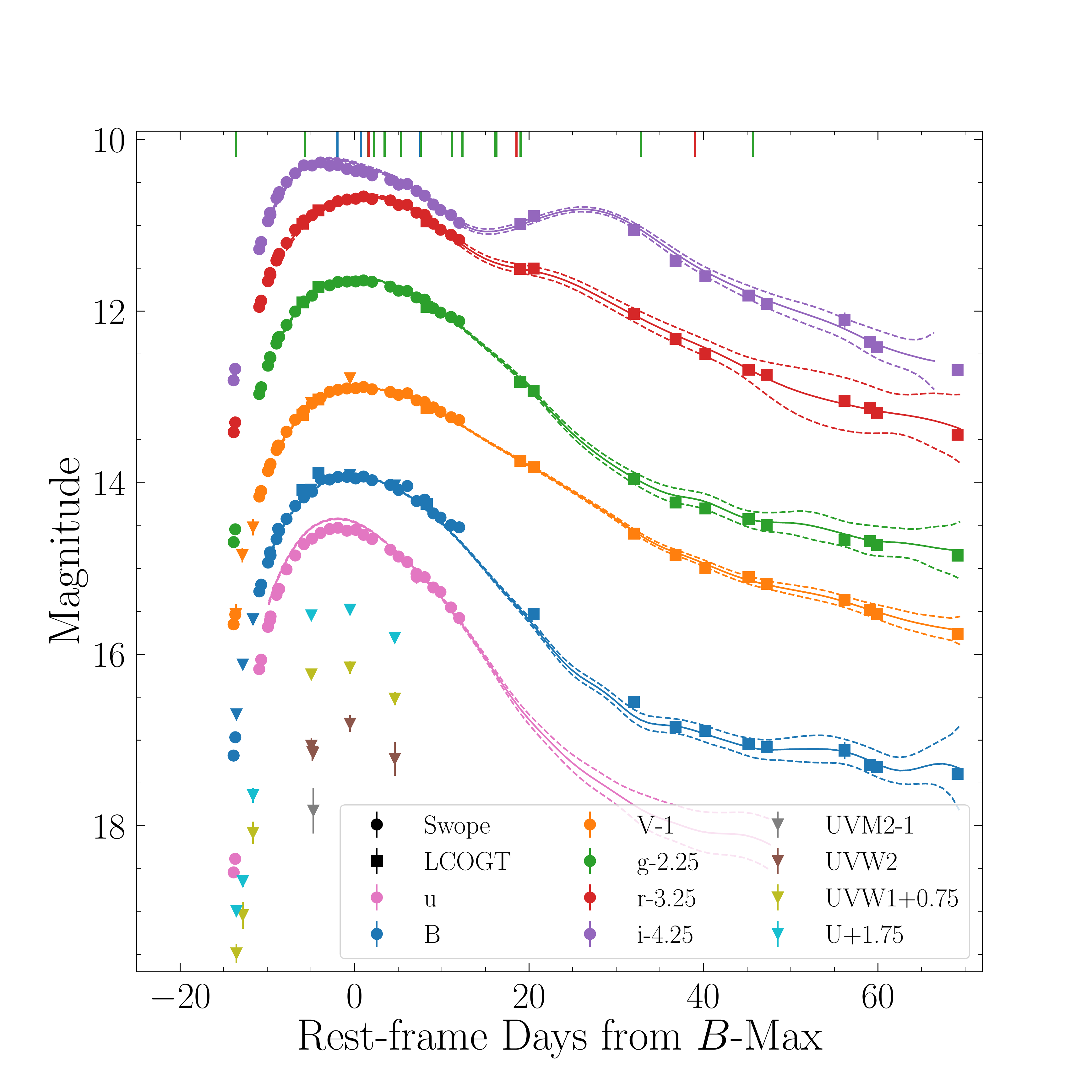}
  \caption{Multi-band light curves of SN~2021fxy. $uBVgri$ data from the Swope 1-m 
  telescope are shown in circles, $BVgri$ data from Las Cumbres Observatory Global 
  Telescope 1-m network in squares, and $UVW2$, $UVM2$, $UVW1$, $U$, and $B$ data 
  from  {\it SWIFT/UVOT} in triangles. The last non-detection from ASAS-SN is noted 
  with an open diamond. Rest-frame epochs of UV (blue), optical (green), and NIR (red) 
  spectra are marked along the top axis. \texttt{SNooPy} fits to the $uBVgri$ photometry are 
  plotted (solid lines), with 1-$\sigma$ errors (dashed lines).}
  \label{fig:lc}
\end{figure*}

The full multi-band light curves are shown in \autoref{fig:lc}, with the 
$uBVgri$ photometry presented in the CSP natural system. Using SuperNovae 
in Object Oriented Python \citep[\texttt{SNooPy},][]{Burns2014}, we fit 
the $uBVgri$ light curves with the \texttt{EBV\_model2}, with the fits 
shown in \autoref{fig:lc}. From the fit, we measure a $B$-band maximum of 
$13.57 \pm 0.01$ mag on $t_{max} = 59305.12 \pm 0.34$ days, corresponding 
to $16.7$ days after last non-detection. The colour-stretch $s_{BV}$ is found 
to be $0.99 \pm 0.03$, which is consistent with a normal-bright SN~Ia. We 
obtain a value of $\Delta m_{15}(B) = 1.05 \pm 0.06 $ mag from the \texttt{SNooPy} 
fit to the multi-band light curves. The distance modulus is estimated from our 
fit as $\mu = 32.86 \pm 0.08$ mag, which is consistent with the value derived 
from the host redshift in \autoref{sec:obs}. The host extinction derived from 
the SN light curves is estimated to be $E(B-V)_{host} = 0.02 \pm 0.06$~mag.

\begin{figure}[t]
  \centering
  \includegraphics[width=\columnwidth]{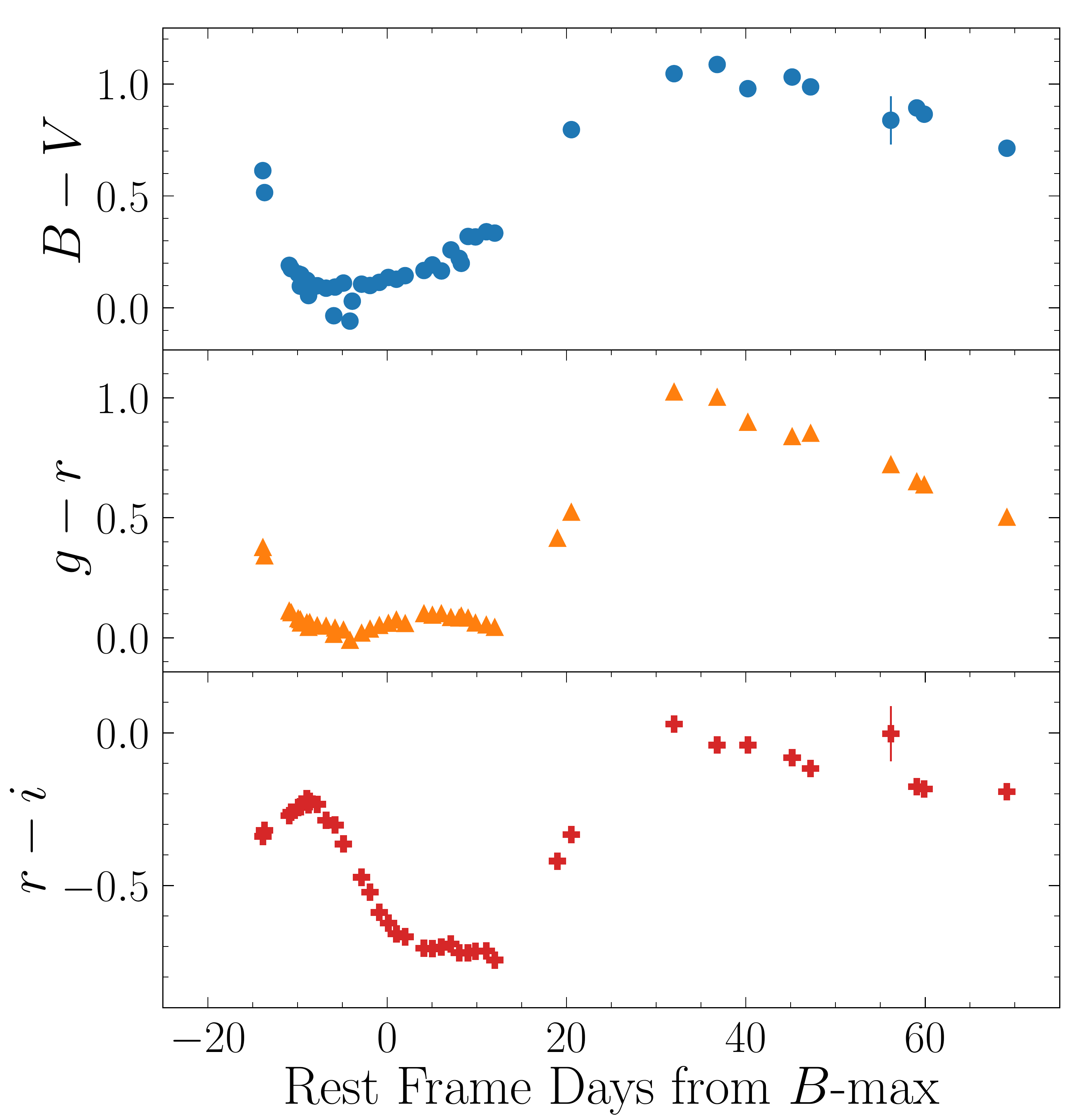}
  \caption{$B-V$, $g-r$, and $r-i$ colour curves of SN~2021fxy in the CSP
  natural system.}
  \label{fig:colour_comp}
\end{figure}

An examination of the early $B-V$ colour evolution, as shown in
\autoref{fig:colour_comp}, reveals that SN~2021fxy follows the ``red'' 
evolutionary track, as defined by \cite{Stritzinger2018}. SN~2021fxy's 
classification as a ``shallow-silicon'' (SS) object within the Branch scheme 
(see Sect.~\ref{sec:spec_analysis}) would make it the first known 
spectroscopically normal SS object to follow the ``red'' track\footnote{
\citet{Stritzinger2018}, \citet{Contreras2018}, and \citet{Cain2018}
describe SN~2012fr as a spectroscopically peculiar SN~Ia lying on the 
border of the CN/SS subgroups, which follows the ``red" evolutionary
track. \citet{Burrow2020} found that SN~2012fr had a $98.5\%$ probability 
of being a SS object.} as all the normal SS/91T-like objects in the 
\cite{Stritzinger2018} sample follow the ``blue" track.

\begin{figure}[t]
  \centering
  \includegraphics[width=\columnwidth]{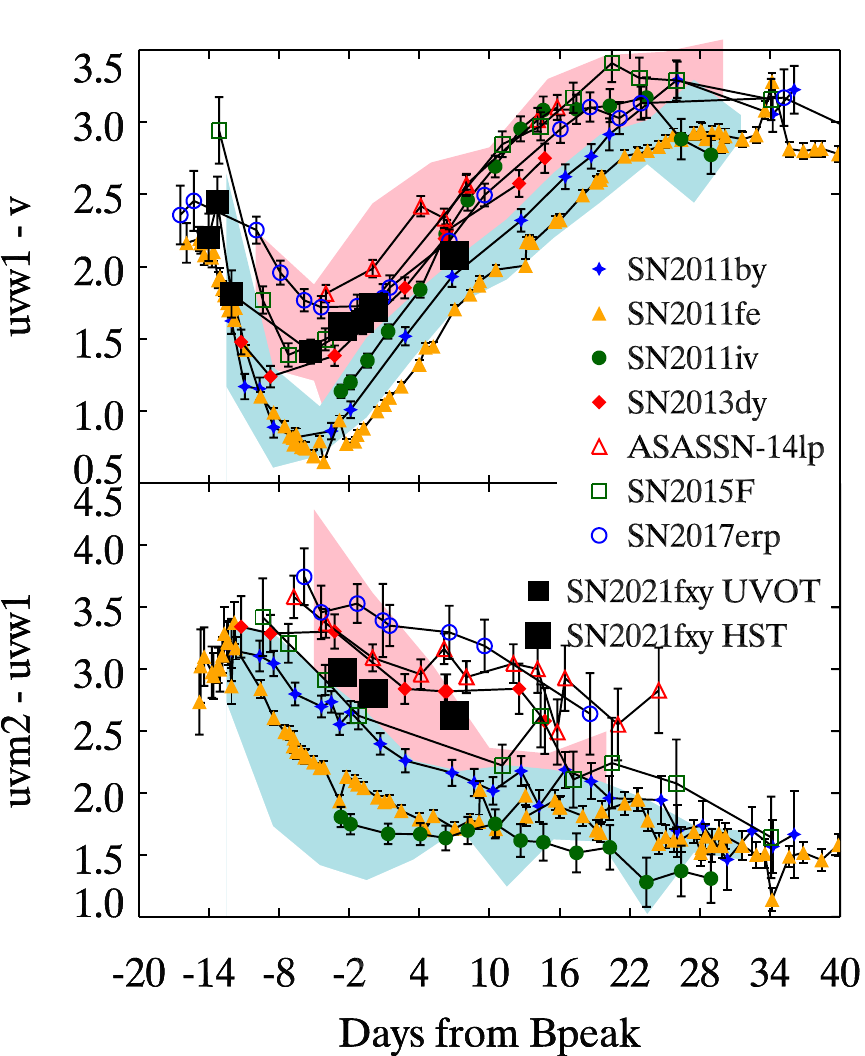}
  \caption{UV colour evolution of SN~2021fxy compared to other spectroscopically
  normal SNe~Ia with {\it HST} UV spectroscopy. The pale blue and red polygons 
  note the locations of NUV-blue and NUV-red objects respectively, as defined in 
  \citet{Milne2013}. The larger black squares are synthetic measurements 
  derived from \textit{HST}/STIS UV spectral observations of SN~2021fxy and 
  the {\it Swift/UVOT} filters \citep{Breeveld2011}.}
  \label{fig:uv_colour_evo}
\end{figure}

The colour evolution in the UV compared to other SNe~Ia with {\it HST} UV 
spectroscopy is shown in \autoref{fig:uv_colour_evo}. To supplement the low 
number of {\it Swift} observations, we compute synthetic photometry using the 
\textit{HST}/STIS spectra and the {\it Swift}/UVOT filters as calibrated by 
\citet{Breeveld2011}. The agreement of {\it Swift}/UVOT photometry and 
spectrophotometry from {\it HST} spectra has been found to agree at the 0.05 mag 
level \citep{Brown_2014}. SN~2021fxy is found to be an NUV-red object in 
the \cite{Milne2013} scheme, and shows evolution similar to other NUV-red 
objects SNe~2013dy, ASASSN-14lp, and 2017erp.

\subsection{Spectroscopic Analysis} \label{sec:spec_analysis}

\begin{figure*}
  \centering
  \includegraphics[width=\textwidth]{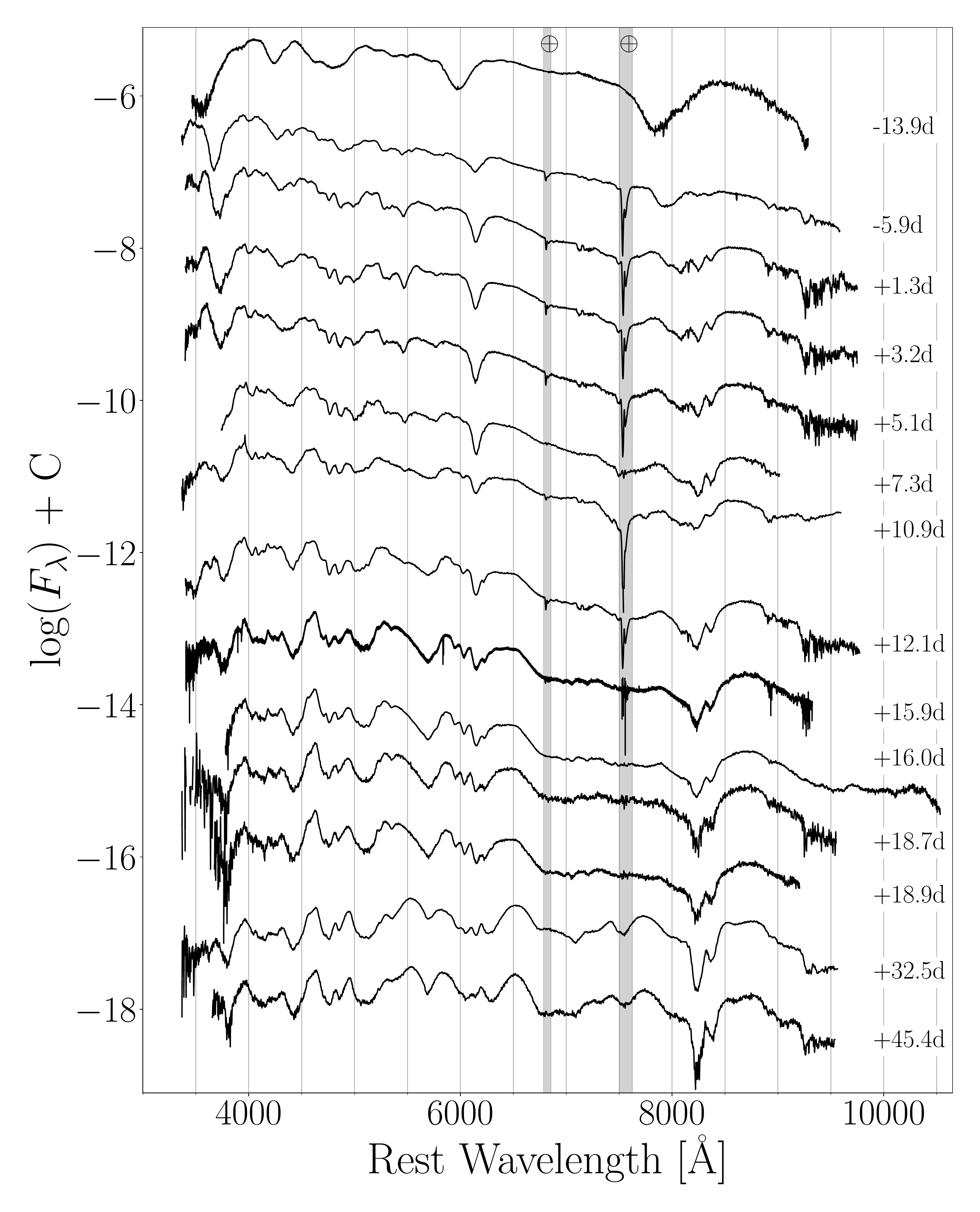}
  \caption{Optical spectra of SN~2021fxy corrected for Milky Way extinction.
    Epoch relative to rest-frame $B$-band maximum are shown next to each 
    spectrum. Gray boxes mark regions of strong telluric absorption.}
  \label{fig:opt_spec}
\end{figure*}

\subsubsection{Optical Spectra}

\begin{figure*}
  \centering
  \includegraphics[width=\textwidth]{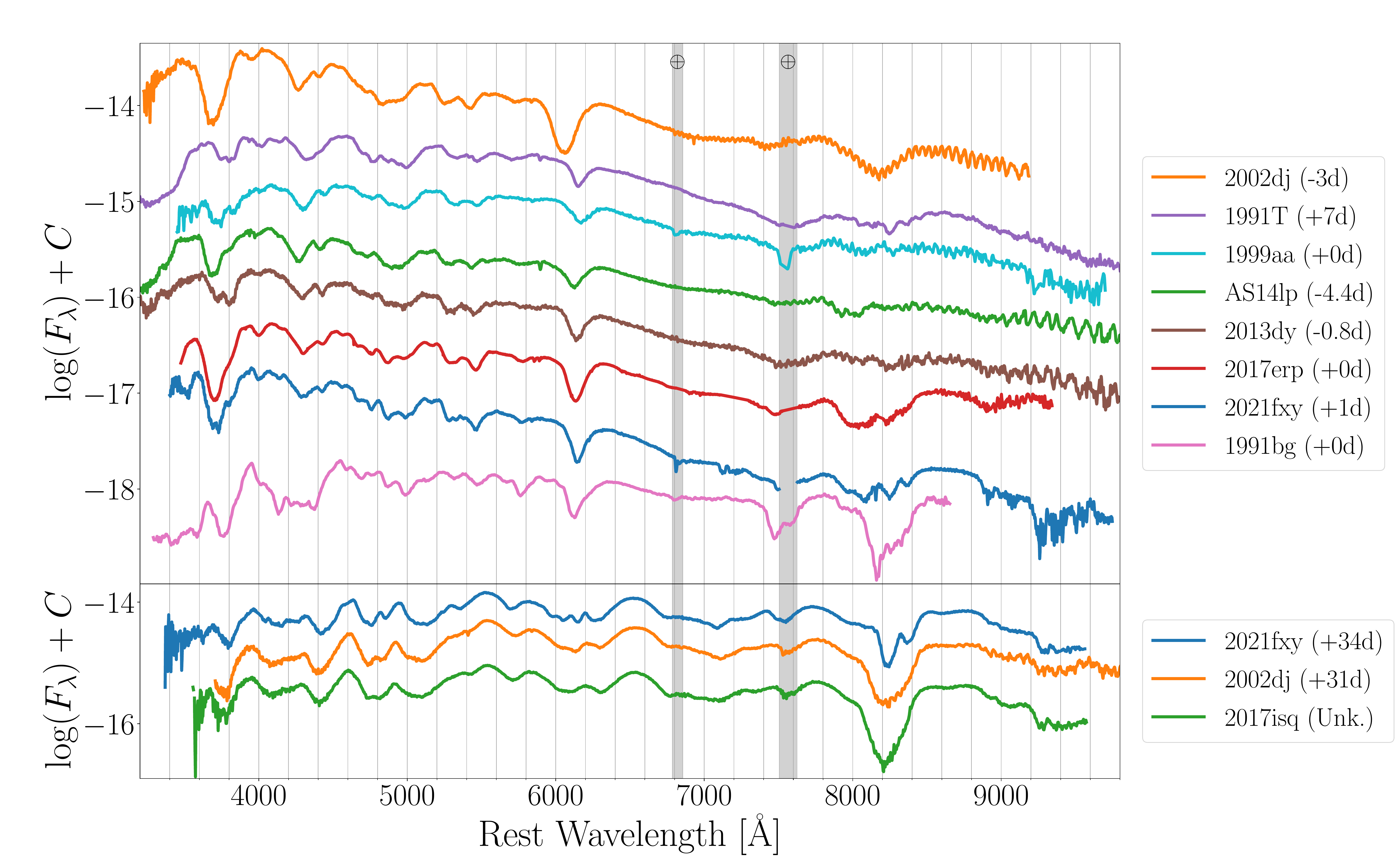}
  \caption{\textit{Top Panel}: Comparison of SN~2021fxy near maximum light to 
  its sibling SN~2002dj (BL), SNe~2013dy and 2017erp (CN), SNe~1991T, 1999aa, 
  and ASASSN-14lp (SS), and SN~1991bg (CL). \textit{Bottom Panel}: Comparison of 
  siblings SNe~2002dj and 2021fxy to potential sibling SN~2017isq roughly one 
  month after maximum light.} 
  \label{fig:host_compare}
\end{figure*}

The optical spectral sequence of SN~2021fxy is shown in \autoref{fig:opt_spec}.
The earliest optical spectrum at $-13.9$ days relative to $B$-band maximum
shows both high velocity \ion{Si}{ii} $\lambda6355$ and \ion{Ca}{ii} NIR 
triplet features at $-18,200$ and $-27,400$ km s$^{-1}$ respectively, as 
measured from the minimum of the absorption troughs. 
Using the blue edge of the absorption troughs as a measure of the maximal 
velocity extent of the line forming ejecta, we estimate the \ion{Si}{ii} 
$\lambda6355$ line extends to at least $-28,000$~\kmps, while the 
\ion{Ca}{ii} extends to at least $-40,000$~\kmps; although strong 
telluric features and possible blending make identification of this edge 
difficult. At $-5.9$ days, the high velocity \ion{Si}{ii} has mostly faded, 
but can still be detected to roughly $-21,000$~\kmps. Meanwhile the 
high velocity \ion{Ca}{ii} remains prominent, particularly in the NIR triplet, 
until roughly $+12$ days, before fully disappearing around $+19$ days. In 
the $-5.9$ day spectrum, material in both the H\&K lines and the NIR triplet 
extend to $-29,000$~\kmps. The \ion{Si}{ii} $\lambda6355$ line has a noticeably 
flat emission peak, indicating that the Si is detached from the photosphere 
\citep{Jeffery1990}. The spectra otherwise resemble that of a typical 
``Branch-normal'' SN~Ia, as shown in \autoref{fig:host_compare}.

\begin{figure}
  \centering
  \includegraphics[trim=1.5cm 0cm 3cm 0cm,clip=True,width=\columnwidth]{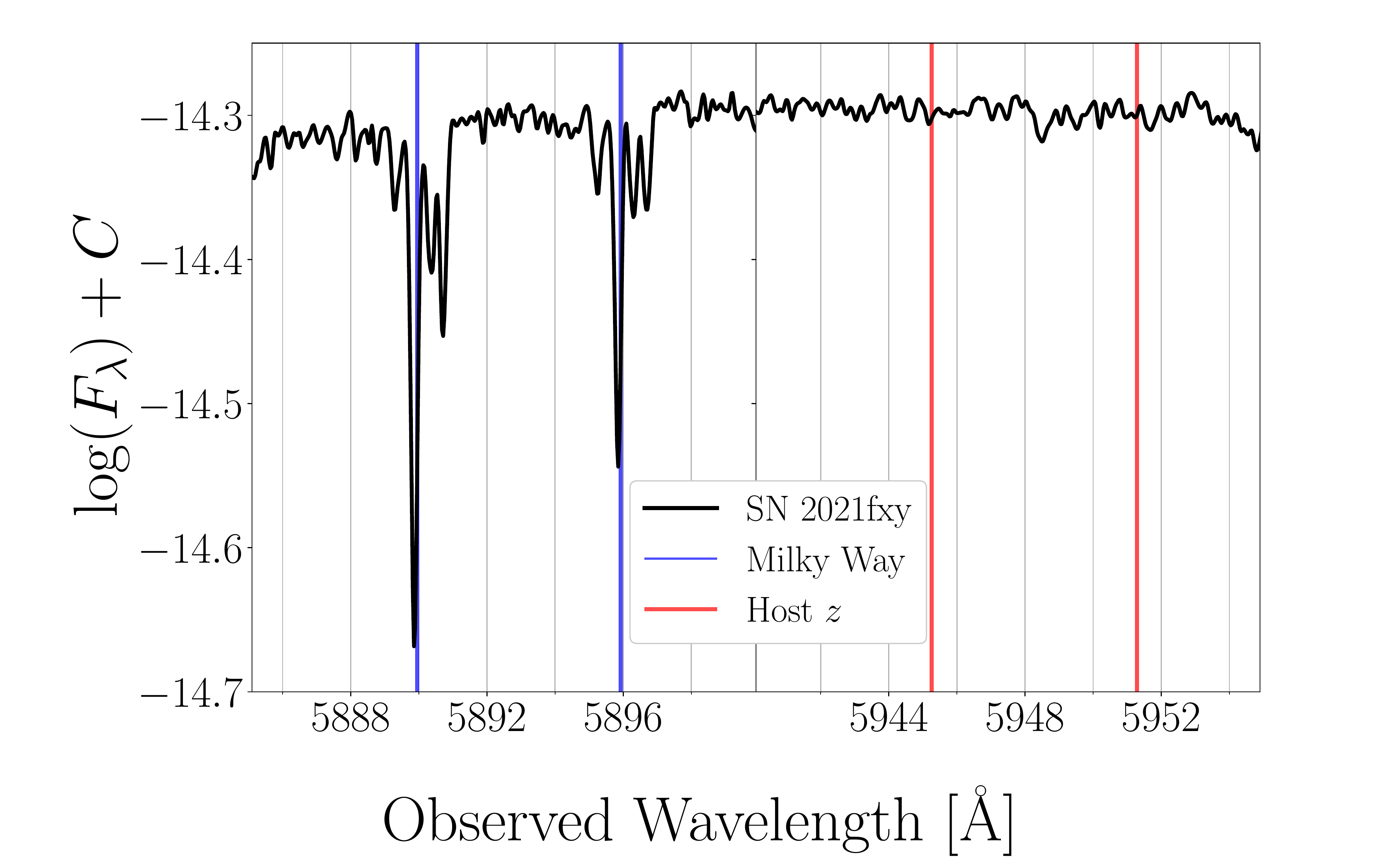}
  \caption{High-resolution spectrum of SN~2021fxy at $+15.9$ days focused 
  on the narrow \ion{Na}{i} D doublets. Four distinct sets of doublets 
  are seen corresponding to interstellar clouds within the Milky Way (left). 
  No detectable \ion{Na}{i} D lines are seen at or near the host redshift 
  of $z=0.0094$.}
  \label{fig:na1d_lines}
\end{figure}

A high-resolution spectra at $+15.9$ days taken with the MIKE spectrograph 
on the Landon T. Clay (Magellan) telescope reveals four distinct 
\ion{Na}{i}~D doublets along the line of sight to the supernova in the 
Milky Way (see \autoref{fig:na1d_lines}). The pseudo-equivalent width of 
the Galactic \ion{Na}{i}~D lines is $0.524 \pm 0.002$~\AA, which implies 
an extinction of $E(B-V)_{MW} = 0.058 \pm 0.039$ mag according to Eq.~(9) of 
\cite{Poznanski2012}, compared to the $E(B-V)_{MW} = 0.084$ mag derived from 
the \citet{Schlafly2011} re-calibration of the \citet{Schlegel1998} dust 
maps, assuming $R_V = 3.1$. Note that, \citet{Phillips:2013} found
that \citet{Poznanski2012} had underestimated their errors by about a
factor of three, thus the two values are consistent with each other.
No absorption from \ion{Na}{i} D is seen 
at or near the redshift of NGC~5018, implying that there is negligible 
host reddening of SN~2021fxy \citep{Phillips:2013}; consistent with a host
galaxy of type E3 and the estimate derived from the SN photometry.

\begin{figure*}
  \centering
  \includegraphics[trim=6cm 0cm 6cm 6cm,clip=True,width=\textwidth]{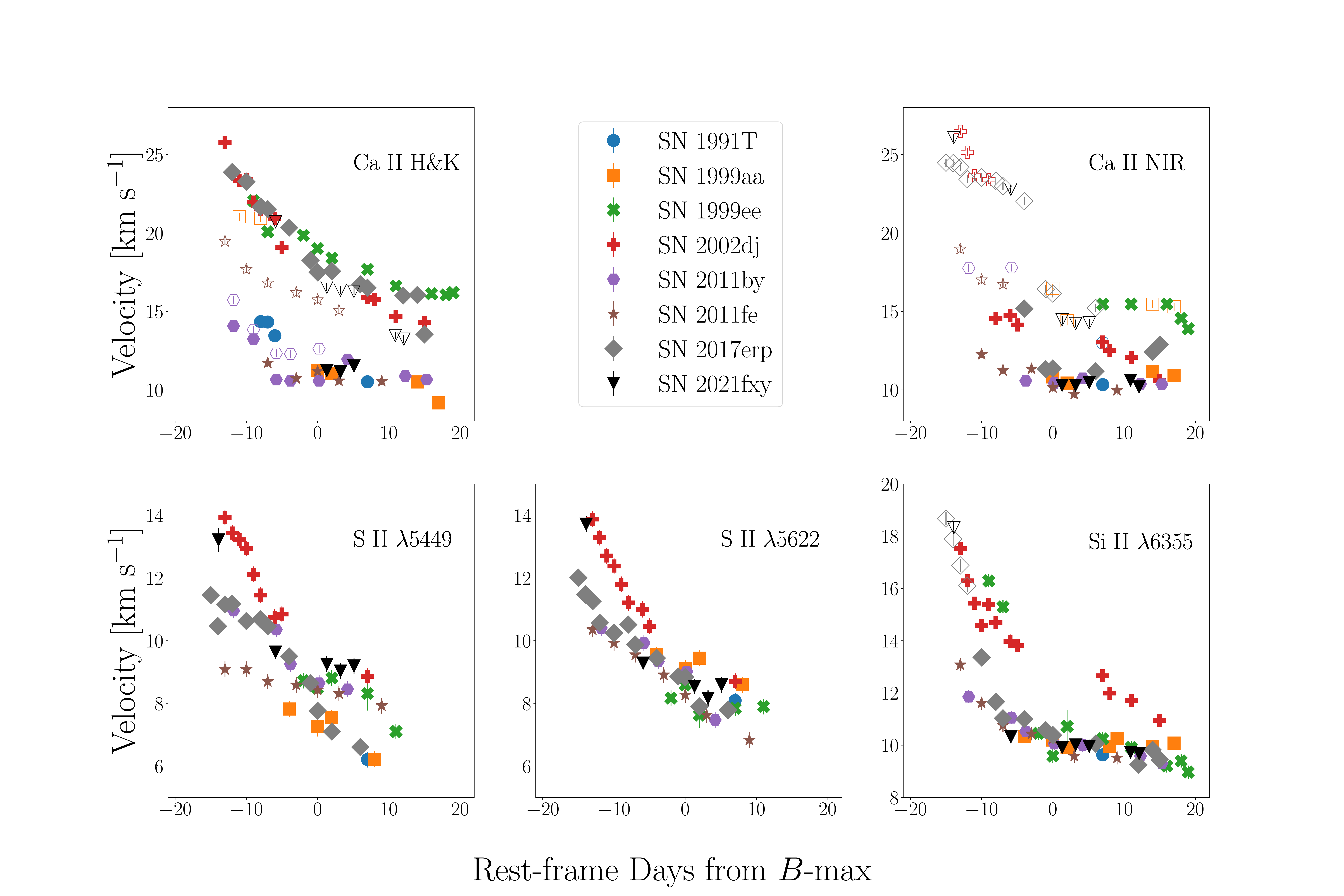}
  \caption{Velocities of prominent SNe~Ia features during the photospheric 
    phase. High-velocity components (when not blended with photospheric components)
    are noted with open symbols. In most cases, the error bars are smaller than 
    the data markers.}
  \label{fig:vels}
\end{figure*}

Velocity measurements of several key SN~Ia features were made
using the Measure Intricate Spectral Features In Transient Spectra
(\texttt{misfits})
package\footnote{\url{https://github.com/sholmbo/misfits}} \citep{Holmbo2020}
and are shown in \autoref{fig:vels}. Fitted spectra are
first smoothed using the Fast Fourier Transform low-pass filter method 
described in \cite{Marion2009}. Next, a raw error spectrum is calculated 
from the difference between the unsmoothed and smoothed spectra, before 
it is smoothed using a Gaussian kernel. This ensure that the smoothed 
errors encompass $68\%$ of the absolute value of the raw error spectrum. 
Measurements of the feature minima are then made by fitting the minimum 
value of the smoothed spectrum over a user-defined range. A Monte Carlo (MC)  
method is applied to generate a new instance of the smoothed spectrum 
and repeat this measurement 1000 times, with the overall error determined 
by adding the measurement error (as defined by the $1\sigma$ spread 
from the MC sample) to the error derived from the instrumental 
resolution (assumed to be 6 \AA\ when not provided) in quadrature. As 
shown in \autoref{fig:vels}, these measurements reveal that SN~2021fxy has 
similar velocities to other well-observed SNe~Ia, including the extreme SS 
object SN~1991T \citep{Filippenko1992,Jeffery1992,Phillips1992}; SS objects 
like SNe~1999aa \citep{Garavini2004,Silverman2012} and 1999ee 
\citep{Hamuy2002,Silverman2012}; and CN objects including SNe~2011by 
\citep{Stahl2020}, 2011fe \citep{Pereira2013,Stahl2020}, and 2017erp 
\citep{Brown2019}. SN~2021fxy's sibling, SN~2002dj (an 2002bo-like or BL) 
is also shown for comparison \citep{Pignata2008}. 

From the measurements in \autoref{fig:vels}, we find that SN~2021fxy is a
``Normal" SN Ia within the Wang scheme, with $v_{Si~II} = 9900 \pm 250$ \kmps
in the $+1.3$ day spectrum. SN~2021fxy belongs to the ``LVG" group in the Benetti 
scheme with a velocity gradient of $\dot{v} = -26 \pm 1$ \kmps.

\begin{figure}
  \centering
  \includegraphics[width=\columnwidth]{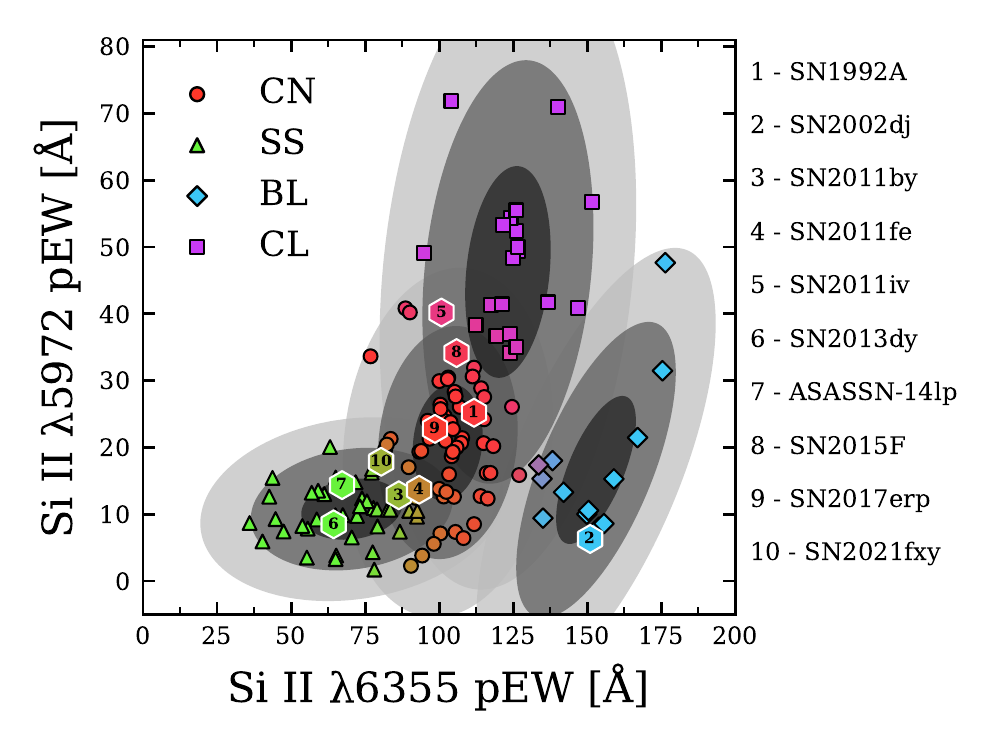}
  \caption{Branch Diagram from \citet{Burrow2020} with SNe~2021fxy, 2002dj, and
    other members of the UV sample of spectroscopically normal SNe~Ia over-plotted. 
    The colouring of each point corresponds to the likelihood of membership in a 
    particular subgroup.}
  \label{fig:branch_diagram}
\end{figure}

Using the Spextractor\footnote{\url{https://github.com/anthonyburrow/spextractor}}
and SNIaDCA\footnote{\url{https://github.com/anthonyburrow/SNIaDCA}}
packages of \cite{Burrow2020}, we measure the pEW's of the \ion{Si}{ii}
$\lambda5972$ and $\lambda6355$ features in the $+1.3$ d spectra to classify 
SN~2021fxy within the Branch scheme \citep{Branch2006}. As shown in
\autoref{fig:branch_diagram}, SN~2021fxy falls near the intersection of
the $2\sigma$ confidence regions of the SS and CN groups, and has as a
$64.1\%$ chance of belonging to the SS subgroup, compared to a $35.7\%$
chance of belonging to the CN subgroup. We note that some of the SNe~Ia 
in the \cite{Burrow2020} sample located within a few \AA{} of SN~2021fxy 
in the Branch diagram change their group membership from CN to SS when 
additional information like the \ion{Si}{ii} velocity at max light and 
maximum $B$-band magnitude are included in the Gaussian Mixture Model
which determines membership in the different Branch subgroups. 
Interestingly, SNe~Ia such as SNe~2011by and 2011fe, which are typically
associated with the core-normals also fall along the border of the CN
and SS subgroups. While differences in the spectral epoch, methodology 
of measuring the pEW's and locations of the boundaries between the different
subgroups may vary from diagram to diagram, it is clear that a significant 
number of ``normal" SNe~Ia lie along the CN/SS border. While SN~1991T-like 
objects have been found to be an extreme subset of the SS group 
\citep{Phillips2022}, further study of the SNe~Ia that reside 
along the CN/SS boundary of the Branch diagram may reveal information about the 
underlying physical differences between the two subgroups. Clearly, SNe~2011by, 
2011fe, and 2021fxy are not SN~1991T-like objects, but rather belong to
a group of objects that are near the middle of the SS/CN continuum. 

\subsubsection{UV Spectra}

\begin{figure*}
  \centering
  \includegraphics[width=\textwidth]{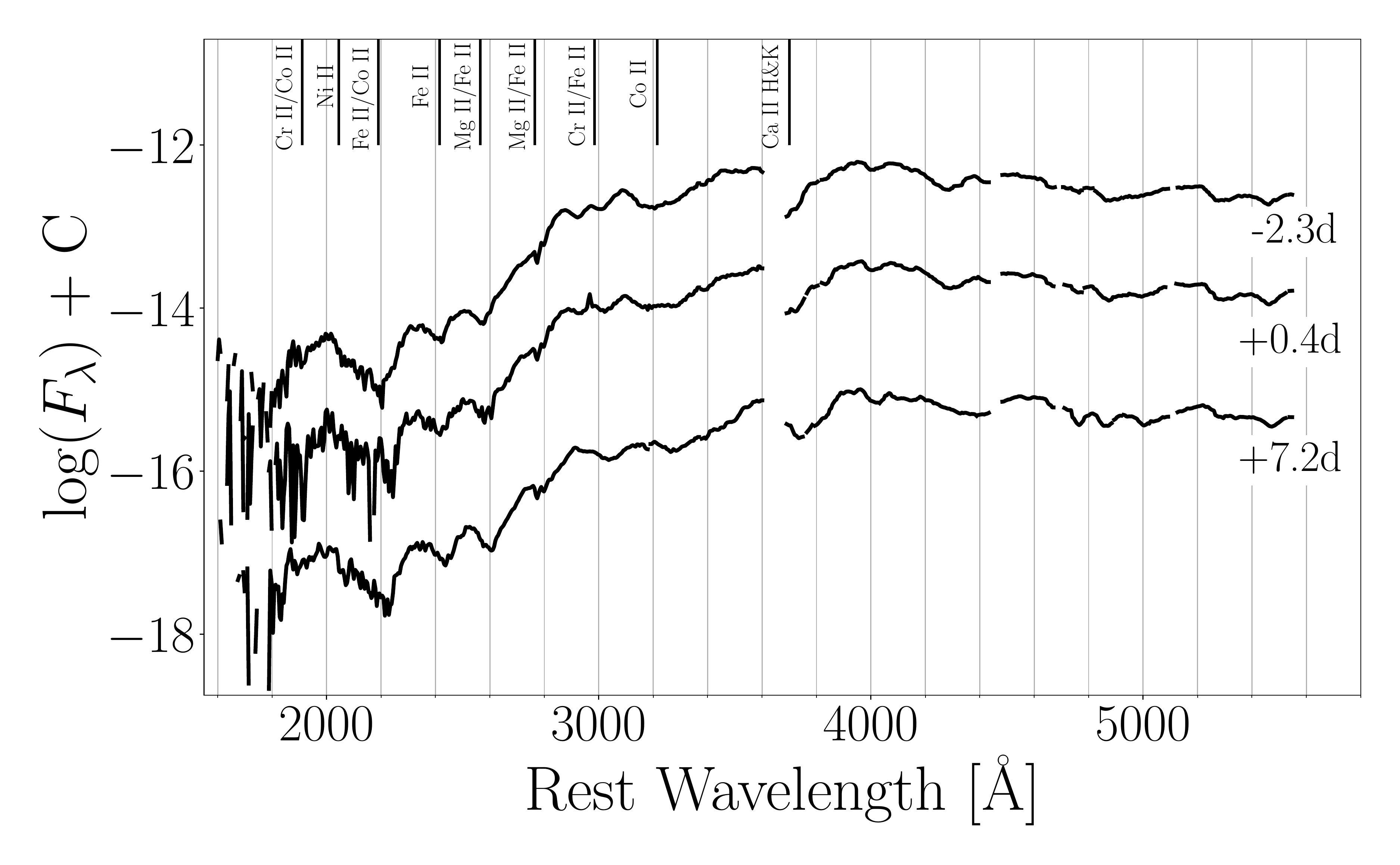}
  \caption{\textit{HST}/STIS spectroscopy of SN~2021fxy corrected for Milky 
    Way extinction. UV line identifications from the near-max models of 
    \citet{DerKacy2020} are shown. Small gaps in the spectra result from 
    the masking of bad pixels during the reduction.}
  \label{fig:uv_spec}
\end{figure*}

The UV spectral sequence obtained with \textit{HST}/STIS is shown in
\autoref{fig:uv_spec}. The UV spectral features are relatively stable 
across the $\sim$10 day interval, with the most prominent evolution 
being the gradual weakening of the \ion{Cr}{ii}/\ion{Co}{ii}/\ion{Fe}{ii} 
blends near $\sim2950$~\AA~and $\sim3200$~\AA. 

\begin{figure*}
  \centering
  \includegraphics[trim=2cm 0cm 6cm 0cm,clip=True,width=\textwidth]{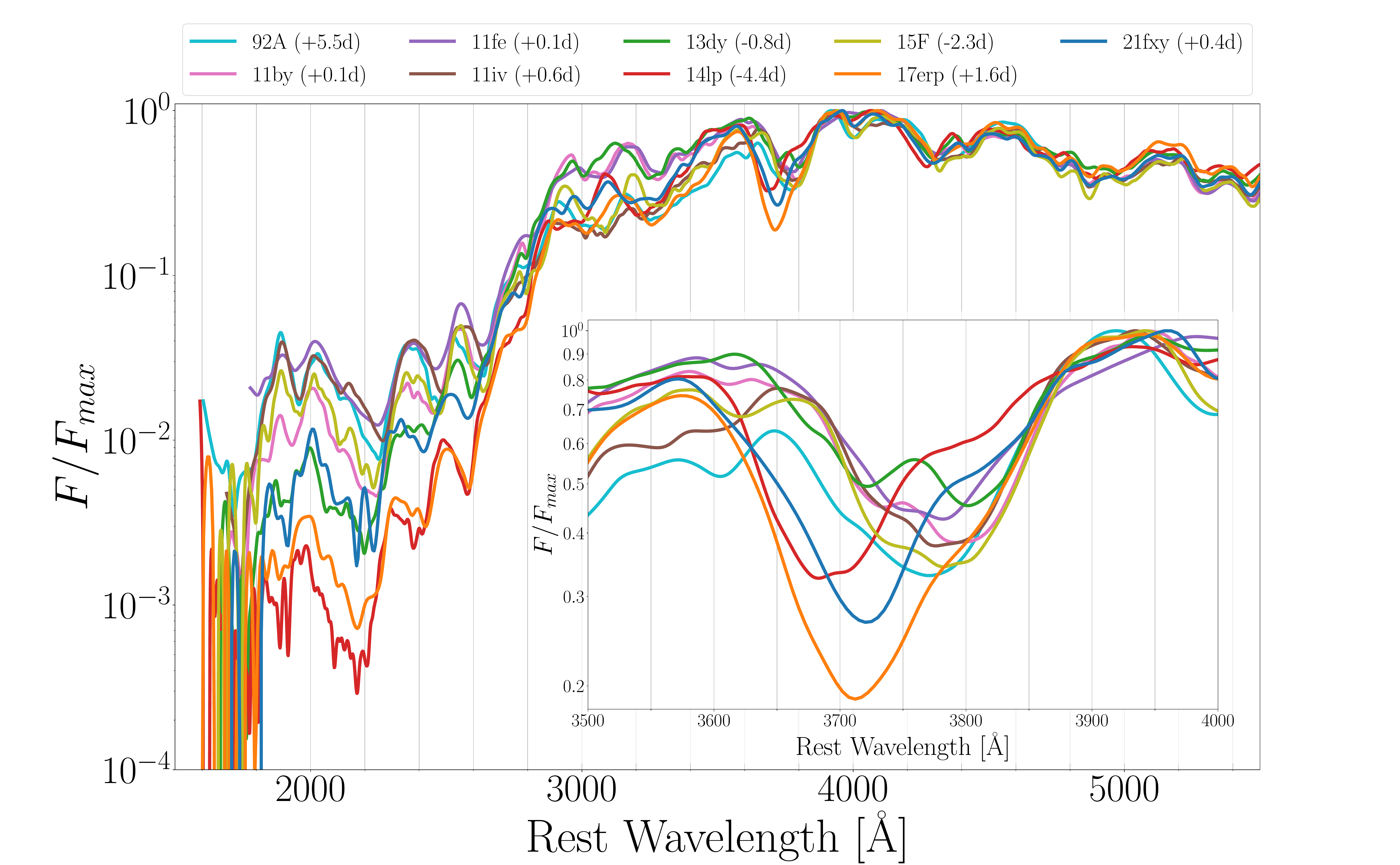}
  \caption{Comparison of SN~2021fxy with other spectroscopically normal SNe~Ia 
    having \textit{HST}/STIS spectra, corrected for Milky Way extinction. Spectra 
    are normalized relative to the maximum flux, which is located at either the 
    \ion{Ca}{ii} H\&K emission peak or the \ion{Si}{ii} blend emission peak at 
    $\sim 4075$~\AA. Along with SNe~2013dy, ASASSN-14lp, and 2017erp, SN~2021fxy 
    shows suppressed flux in the mid-UV and blue-shifted mid-UV features relative 
    to SNe~Ia without suppressed mid-UV fluxes. {\it Inset:} Close up of the 
    \ion{Ca}{ii} H\&K features, showing that SNe~Ia with MUV flux suppression have
    strong HV \ion{Ca}{ii} H\&K components which dominate the feature.}
  \label{fig:muv_flux_mwcor}
\end{figure*}

\autoref{fig:muv_flux_mwcor} shows the maximum light UV spectrum of 
SN~2021fxy plotted against the spectra of other spectroscopically normal 
SNe~Ia with {\it HST} UV spectra close to maximum light, including 
SNe~2011by \citep{Foley2013,Graham2015}, 
2011fe \citep{Pereira2013,Mazzali2014}, 2011iv \citep{Foley2012,Gall2018}, 
2013dy \citep{Pan2015}, ASASSN-14lp \citep{Shappee2016,Foley2016}, 
2015F \citep{Foley2016,Burns2018}, and 2017erp \citep{Brown2019}.
From this sample, we see that four objects, SNe~2013dy, ASSASN-14lp, 2017erp, 
and 2021fxy show ``suppressed" flux in the mid-UV relative to the other 
SNe~Ia in the sample, which show less variation in their relative fluxes 
throughout the mid-UV. These four suppressed SNe~also have mid-UV 
features which are blue-shifted relative to their un-suppressed 
counterparts. Both the ``suppressed" and ``un-suppressed" subsets show 
no common behaviors in either their near-UV spectral features or flux 
levels. However, the two subsets do show differences in their \ion{Ca}{ii} 
H\&K features, with the suppressed SNe~Ia possessing strong high velocity 
(HV) components that dominate the H\&K feature. Unsuppressed SNe~Ia show 
much weaker HV components and are dominated by the photospheric component. 
In the cases of the SNe~Ia with {\it HST} spectra extending to the red half of 
the optical and NIR, those with MUV suppression (SN~2013dy and ASASSN-14lp) 
show higher flux levels than the unsuppressed SNe~Ia, although it is unclear 
how much of these flux differences may be due to variations in host reddening. 
The source of this MUV suppression and its connection to feature locations 
and the \ion{Ca}{ii} H\&K lines is explored further in Sect.~\ref{sec:muv_sup}.

\subsubsection{NIR Spectra}

\begin{figure*}
    \centering
    \includegraphics[width=\textwidth]{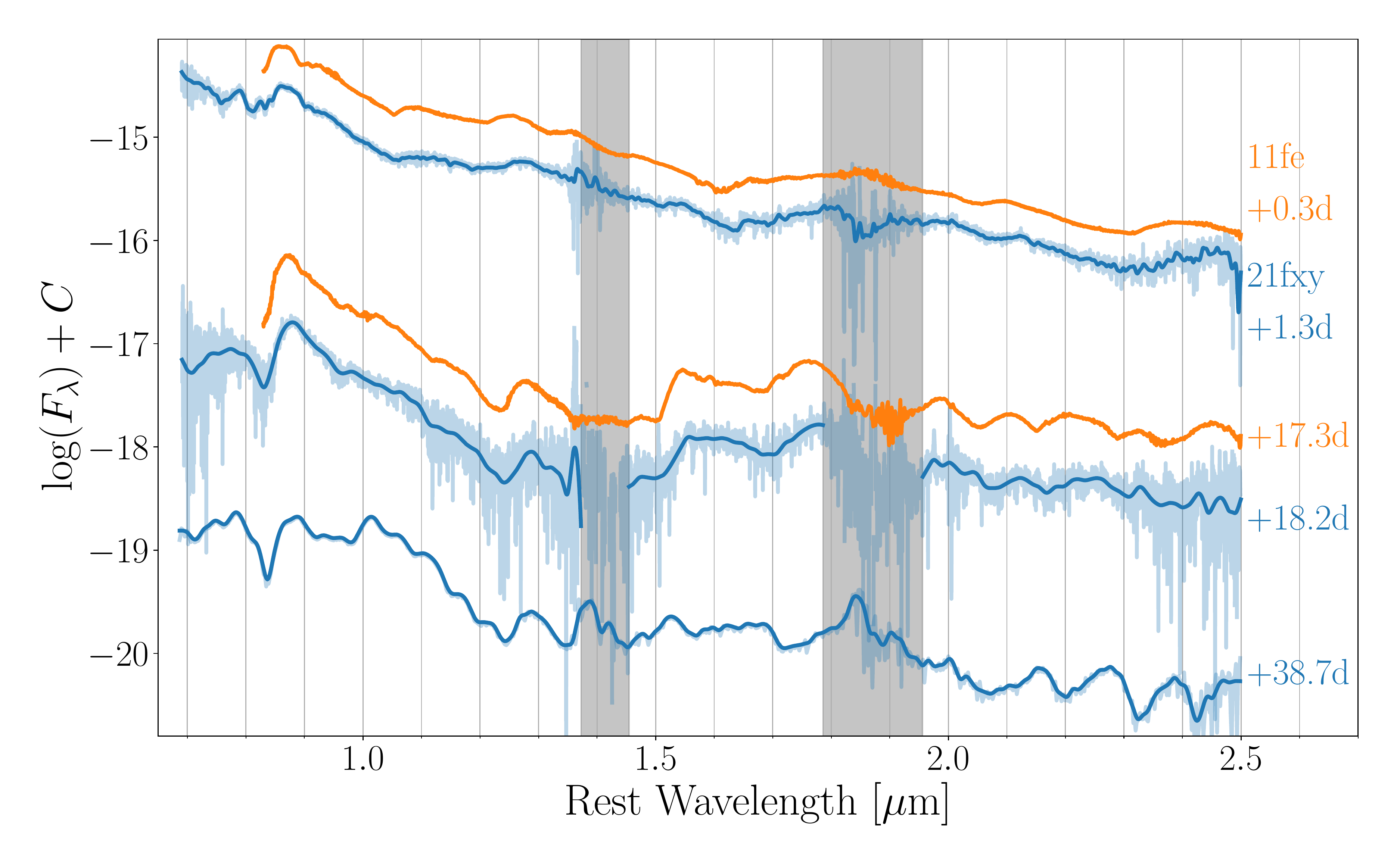}
    \caption{NIR spectra of SN~2021fxy. NIR spectra of SN~2011fe from \citet{Hsiao2013} 
    at similar epochs are shown for comparison. Smoothed spectra are plotted in the 
    foreground, with the raw spectra shown behind. The epoch relative to rest-frame
    $B$-band maximum is noted next to each spectrum. Grey boxes mark regions of strong 
    telluric absorption. The noisy telluric regions between the NIR bands in the 
    $+18.2$ day smoothed spectrum are removed for clarity.}
    \label{fig:nir_spec}
\end{figure*}

NIR spectra of SN~2021fxy are shown in \autoref{fig:nir_spec}. Compared to 
spectra of SN~2011fe at similar epochs \citep{Hsiao2013}, the $+1.3$ day and 
$+18.2$ day spectra show broad similarities. Near maximum light, the 
\ion{Mg}{i} and \ion{Mg}{ii} features between 0.9 $\mu$m and 1.1 $\mu$m appear 
significantly weaker in SN~2021fxy. A close examination of the \ion{C}{i} 
$\lambda1.0693$ and \ion{Mg}{ii} $\lambda1.0927$ blend reveals that the feature 
is so weak that relative to the noise in the spectrum, we cannot conclusively 
identify the presence of either line. In the $+18.2$ day spectrum, high noise 
levels make identifications of many weak features difficult. The noise also 
complicates the measurement of the properties of the $H$-band break, $v_{edge}$. 
Using the same procedure as \citet{Ashall2019a,Ashall2019b} we measure 
$v_{edge} = 14,100 \pm 100$  km~s$^{-1}$, which is slightly higher than other 
measurements of $v_{edge}$ at that epoch in the Ashall sample for SNe~Ia
with $s_{BV} \gtrsim 1$, including SN~2011fe. We note that due to a bump in 
the  spectra on the blueward side of the $H$-band break, an additional 
smoothing step using a 1-dimensional Gaussian kernel was required prior 
beginning the Guassian fit in order to achieve convergence. As a result, the 
reported error of $v_{edge}$ accurately captures the statistical error 
associated with the fitting of the Gaussian center, but underestimates the 
total error. Furthermore, some SNe~Ia in the CSP-II NIR spectral sample show 
blending on the blue side of the $H$-band break by an unidentified feature
around $+20$ days, which complicates the the fit 
\citep{Hsiao2019,Ashall2019a,Ashall2019b}.
The emission peaks from \ion{Co}{ii} lines and blends of Iron Group Elements 
(IGEs) appear noticeably slower than those in 2011fe across all bands. 
By $+38$ days the spectra is dominated by IGE lines, similar to other NIR 
spectra at similar epochs in the sample of \citet{Marion2009}.

\section{Discussion}\label{sec:discussion}

From \autoref{fig:muv_flux_mwcor}, it is clear that MUV suppression is 
a common feature of SNe~Ia; and that the strength of this suppression
varies from object to object. However, explaining the root causes of 
these variations is difficult, as spectral formation in the ultraviolet 
is extremely complicated. Several physical factors, such as 
metallicity, density structure, and luminosity are known to have 
strong impacts on the observed spectra. All of these factors 
are inter-related, making the identification of which variable (or 
combination of variables) are responsible for the observed differences 
between SNe~Ia quite difficult. Host reddening also becomes significant 
in the UV, as small differences in the estimate of the host $R_{V}$ can 
significantly alter the observed flux levels. Therefore, rather than 
attempt to disentangle these related effects, we investigate the impact
of these various factors, on-by-one, in order to determine how much
of the observed MUV suppression they can explain.

\subsection{Mid-UV Suppression from Host Extinction?} \label{sec:muv_sup}

\begin{figure*}
  \centering
  \includegraphics[width=\textwidth]{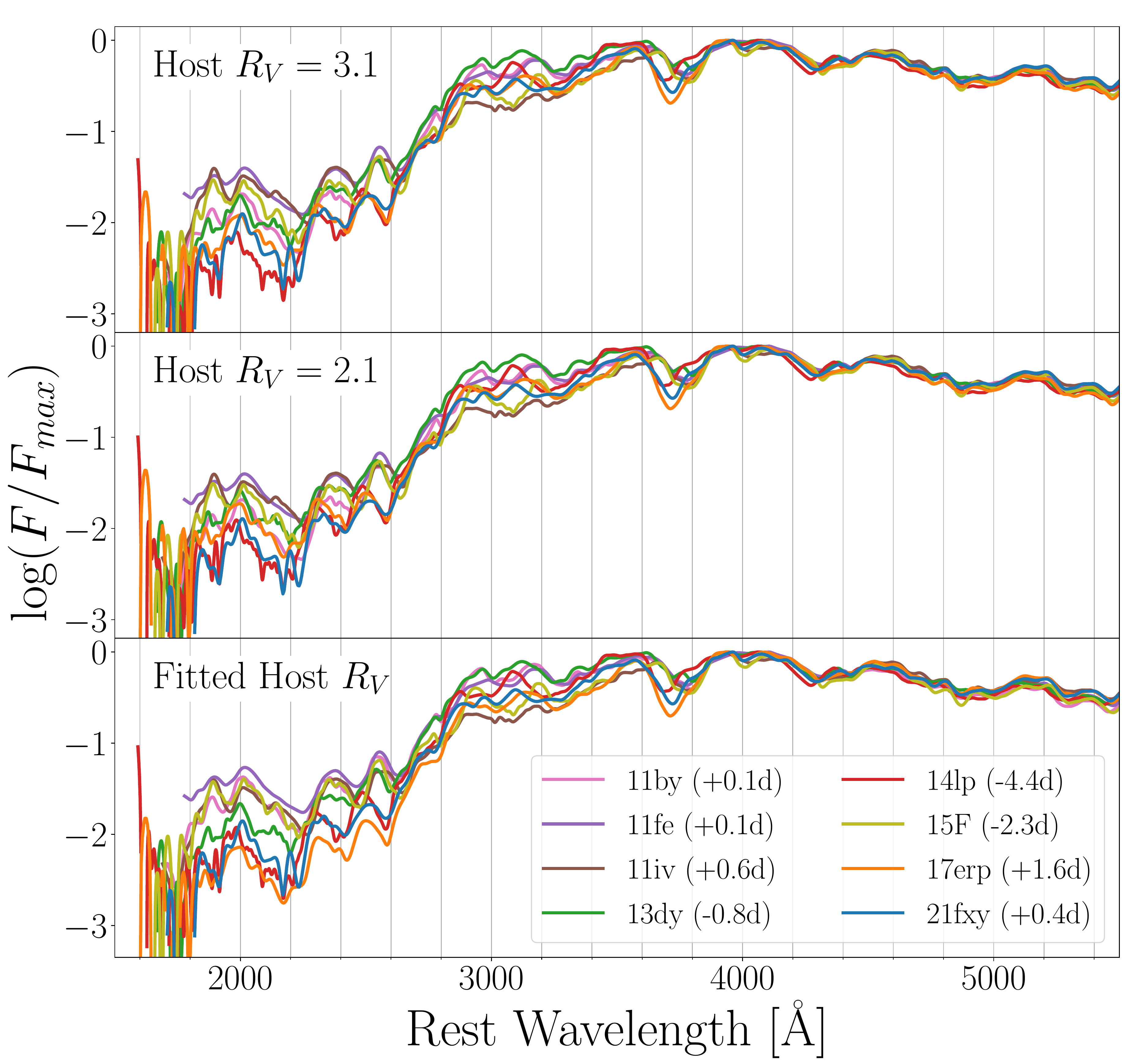}
  \caption{Same as \autoref{fig:muv_flux_mwcor}, but corrected for 
  host extinction assuming $R_V=3.1$ (top), $R_V=2.1$ (middle), and 
  the fitted $R_V$ (bottom). SN~1992A is omitted due to insufficient 
  photometry necessary for robust estimations of $E(B-V)_{host}$ and $R_V$.}
  \label{fig:muv_rv}
\end{figure*}

One potential source of the mid-UV suppression in SNe~Ia is the reddening of 
the SNe by dust in the host galaxy. However, estimating the amount of host 
extinction can be difficult, with different methods yielding significantly 
different results \citep[for an example, see SN~2017erp in][]{Brown2019}. 
Additionally, numerous studies have shown that dust properties vary across 
different galaxies, and their extinction laws have different forms than that 
of the Milky Way
\citep[see, for example,][and references therein]{Mathis:1990,Phillips:2013}.
Therefore, we attempt to correct for the host extinction using the published
values of $E(B-V)_{host}$ in combination with $R_V$ values of ($R_V=3.1$) 
and ($R_V=2.1$), representing hosts with Milky Way like extinction and 
low-metallicity hosts like the Small Magellanic Cloud 
\citep[SMC;][]{Gordon:2014,YanchulovaMerica-Jones:2021}, respectively. We also 
use the supernovae light curves to attempt to fit the value of $R_V$ in the host galaxy using 
\texttt{SNooPy}'s \texttt{colour\_model}, which simultaneously fits both $E(B-V)_{host}$ 
and $R_V$ based on the intrinsic SN~Ia colours determined by \cite{Burns2014}. 
Since both parameters appear within the same fitting term; they are covariant; 
and in cases where the total host extinction is small, the fit can become 
insensitive to the value of $R_V$. If this occurs, we impose a prior derived 
from a Gaussian Mixture Model representing the distribution of $R_V$ as 
determined from the CSP-I sample in \citet{Burns2014}, and the data is refit 
using a Markov Chain Monte Carlo (MCMC) method. The results of these fits are shown 
in \autoref{tab:host_extinction}. UV spectra corrected for this host extinction 
with the fitted values of $R_V$ and $E(B-V)$ are shown in \autoref{fig:muv_rv}. 

No matter which value of $R_V$ is chosen to correct for host extinction, a 
subset of our objects still show significant MUV suppression. Corrections 
for host extinction do, however, reduce the spread in relative flux between 
SNe showing MUV suppression. After correcting for host extinction, SN~2013dy 
no longer shows significant MUV flux suppression, instead showing relative 
fluxes consistent with unsuppressed SNe~Ia. However, SN~2013dy was one of 
multiple SNe (along with SNe~2011fe, 2011iv, and 2021fxy) with \texttt{SNooPy} fits 
that were initially insensitive to the value of $R_V$, but was the only one 
of this group where the MCMC fitting resulted in significant host extinction. 
All SNe that required MCMC fits show no evidence of significant host extinction 
from spectral observations. Therefore, it is likely that the values derived by 
\texttt{SNooPy} represent over-estimates of the host extinction for these four SNe.

\begin{table}
\centering
\caption{Host Extinction Fitting with \texttt{SNooPy} \label{tab:host_extinction}}
\begin{tabular}{ccc}
  \hline
  Object & $E(B-V)_{host}$ [mag] & $R_V$ \\
  \hline
  2011by & $0.19 \pm 0.06$ & $3.17 \pm 0.52$ \\
  2011fe* & $0.08 \pm 0.06$ & $3.10 \pm 0.16$ \\
  2011iv* & $-0.02 \pm 0.06$ & $3.11 \pm 0.06$ \\
  2013dy* & $0.23 \pm 0.06$ & $3.10 \pm 0.23$ \\
  ASASSN-14lp & $0.34 \pm 0.06$ & $2.27 \pm 0.17$ \\
  2015F & $0.15 \pm 0.06$ & $4.09 \pm 0.31$ \\
  2017erp & $0.18 \pm 0.06$ & $2.80 \pm 0.51$ \\
  2021fxy* & $0.05 \pm 0.06$ & $3.11 \pm 0.05$ \\
  \hline 
\end{tabular} \\
*Objects for which the initial fit was insensitive to $R_V$.
\end{table}

\subsection{Common Properties of MUV-suppressed SNe~Ia} \label{sec:17erp_compare}

With host extinction unable to explain the observed MUV suppression
for those objects showing significantly suppressed MUV fluxes, 
we now turn our focus to identifying physical explanations for this 
behavior based on commonalities in the observational properties of 
the SNe in our UV sample. Examining the SNe~Ia showing evidence of 
MUV suppression, there are a few commonalities shared by all four 
objects that are readily apparent. 

\begin{figure*}
    \centering
    \includegraphics[width=\textwidth]{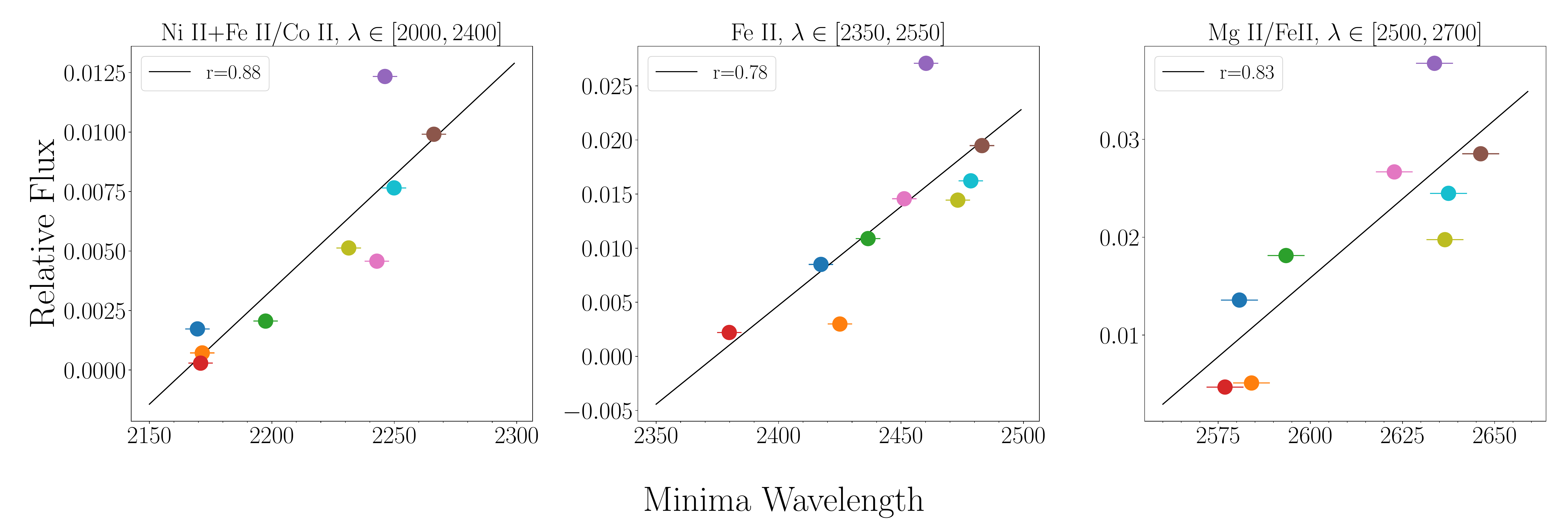}
    \caption{The relative flux ($F/F_{max}$) versus minimum wavelength for three 
    MUV features, with the best-fitting linear regression and Pearson 
    correlation coefficient $r$, also shown. For all data points, the 
    relative flux error bars are smaller than the markers. The spectra 
    were corrected for Milky Way extinction, but not host extinction. 
    Data points share the same colouring as \autoref{fig:muv_flux_mwcor}.} 
    \label{fig:muv_features}
\end{figure*}

As noted in Sect.~\ref{sec:spec_analysis}, those SNe~Ia with MUV flux 
suppression also have feature minima in the mid-UV which are bluer relative 
to SNe~Ia lacking MUV flux suppression. The relationship between the 
location of the flux minima for three mid-UV features (the \ion{Fe}{ii}/\ion{Co}{ii} 
blend between $2000-2400$ \AA, the \ion{Fe}{ii} feature between $2350-2550$ \AA, 
and the \ion{Fe}{ii}/\ion{Mg}{ii} blend between $2500-2700$ \AA) and the 
relative flux at those minima are shown in \autoref{fig:muv_features}. The 
relationship is strongest in the \ion{Fe}{ii}/\ion{Co}{ii} blend, but is 
present in all three features. Similarly, the correlations are strongest
when the spectra are corrected only for MW extinction, as shown in 
\autoref{fig:muv_rv}, but the effect persists for all variations of host 
extinction corrections. This effect arises naturally, as the opacity in the 
mid-UV originates primarily from the line blanketing of iron group elements 
\citep{DerKacy2020}, which increase the effective opacity in the mid-UV as 
their velocity increases \citep{Wang2012}. 

Further support for the presence of high velocity material in MUV-suppressed 
SNe~Ia is found in the obvious dominance of the HV component of the \ion{Ca}{ii} 
H\&K lines over the photospheric components in their spectra. We can quantify 
this dominance using the $R_{HVF}$ measure established by \cite{Childress2014}, 
originally defined for the \ion{Ca}{ii} NIR triplet but now redefined 
here for the H\&K lines as:

\begin{equation}
    R_{HVF} = \frac{pEW(\text{HVF}_{\text{H\&K}})}{pEW(\text{PVF}_{\text{H\&K}})}.
\end{equation}

\begin{figure*}
    \centering
    \includegraphics[width=\textwidth]{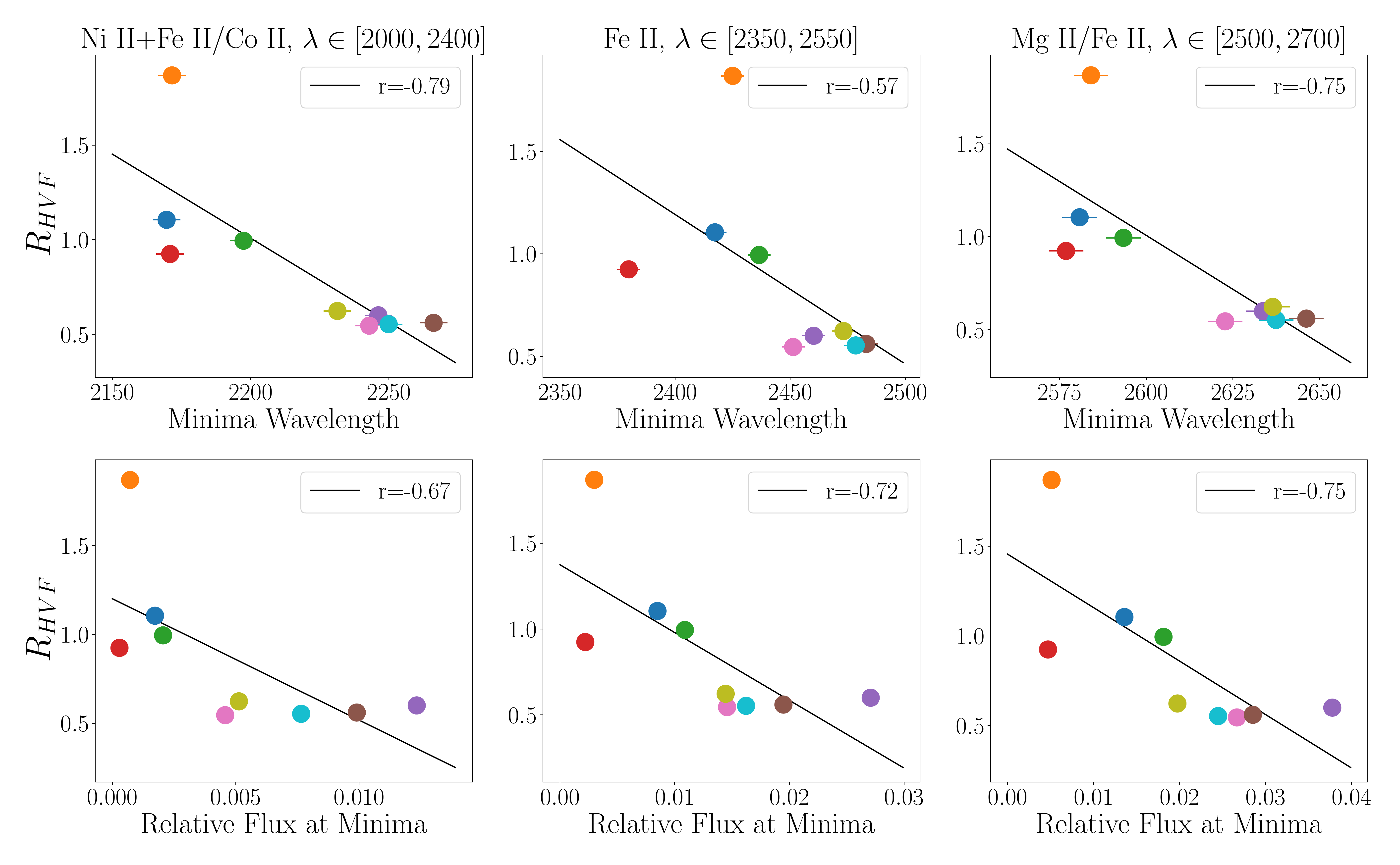}
    \caption{{\it Top Row}: $R_{HVF}$ versus minimum wavelength for 
    three MUV features. The best-fitting linear regression and the 
    Pearson correlation coefficient $r$, are also shown. {\it Bottom Row}: 
    $R_{HVF}$ versus relative flux ($F/F_{max})$ at the minimum wavelength
    for the same MUV features. In all panels, the errors in $R_{HVF}$ are
    smaller than the markers, while in the bottom panels the relative flux error 
    bars are smaller than the markers. As in \autoref{fig:muv_features} spectra 
    corrected for Milky Way extinction, but not host-galaxy extinction. Data 
    points share the same colouring as \autoref{fig:muv_flux_mwcor}.}
    \label{fig:rhvf}
\end{figure*}

We find that MUV-suppressed SNe have values of $R_{HVF} \gtrsim 0.9$, while 
non-suppressed SNe have $R_{HVF} \lesssim 0.6$. $R_{HVF}$ correlates with both 
the minimum wavelength and the relative flux at those minima for the three 
features specified in \autoref{sec:muv_sup}, although it correlates more 
strongly with the minimum wavelengths. As before, the correlations are strongest 
in the \ion{Fe}{ii}/\ion{Co}{ii} blend, but are present in the other two MUV 
features, and are still correlated after correcting for host extinction. These 
correlations are shown in \autoref{fig:rhvf}, with the pEW values shown in 
\autoref{tab:pew}.

\begin{table}
\centering
\caption{pEW Measurements \label{tab:pew}}
\begin{tabular}{cccc}
  \hline
  Object & HVF pEW [\AA] & PVF pEW [\AA] & $R_{HVF}$ \\
  \hline
  1992A & $26.96 \pm 0.95$ & $48.75 \pm 0.62$ & $ 0.55 \pm 0.02$ \\
  2011by & $26.43 \pm 0.24$ & $48.46 \pm 0.25$ & $0.55 \pm 0.01$ \\
  2011fe & $29.29 \pm 0.01$ & $48.81 \pm 0.01$ & $0.60 \pm 0.01$ \\
  2011iv & $23.88 \pm 0.26$ & $42.62 \pm 0.24$ & $0.56 \pm 0.01$ \\
  2013dy & $37.18 \pm 0.26$ & $37.39 \pm 0.26$ & $0.99 \pm 0.01$ \\
  ASASSN-14lp & $58.06 \pm 0.14$ & $62.83 \pm 0.50$ & $0.92 \pm 0.01$ \\
  2015F & $26.38 \pm 0.32$ & $42.33 \pm 0.29$ & $0.62 \pm 0.01$ \\
  2017erp & $66.95 \pm 0.52$ & $35.83 \pm 0.51$ & $1.87 \pm 0.03$ \\
  2021fxy & $68.78 \pm 2.36$ & $62.25 \pm 1.36$ & $1.10 \pm 0.04$ \\
  \hline
\end{tabular}
\end{table}

Other effects may also play a role in producing MUV flux suppression 
coupled with a blue-shift in MUV features, including changes in 
density structure or progenitor metallicity. This same blue-shift in the 
mid-UV features was achieved by \cite{Barna2021} in their modeling of 
ASASSN-14lp by using a modified, shallower version of the W7 density profile. 
Because both shallower density profiles and higher line velocities produce 
broader lines, it is difficult to distinguish between the two effects as the 
primary source of the flux suppression from radiative transfer effects alone.
As the amount of metals in the outer ejecta increases, so does the strength 
of the line blanketing; resulting in the photosphere being located at higher 
velocities, and thus suppressing flux in the MUV. However, large changes in 
metallicity cannot be responsible for this suppression, as they induce other 
changes in the broader UV spectrum that are not seen in the observed spectra 
(see Sect.~\ref{sec:17erp_compare} for more details). \cite{DerKacy2020} 
also found that their lower luminosity models produced these same MUV 
features blue-shifted relative to their higher luminosity models; 
primarily due to temperature variations in the outer portions of the 
ejecta altering the ionization balance between \ion{Fe}{ii} and 
\ion{Fe}{iii} significantly enough to change the relative composition 
of the line blends that comprise these MUV features. If the MUV
suppression is due to higher velocities exhibited in observed
spectral features, one would expect that BL or High-Velocity SNe~Ia
in general would exhibit a strong MUV suppression.

Turning our attention to properties beyond those in the UV spectra,
in \autoref{fig:muv_flux_mwcor}, we see that in the mid-UV, SNe~2013dy 
and 2021fxy have similar flux values and feature locations. The same is 
true of ASASSN-14lp and SN~2017erp. However, in the near-UV, SN~2013dy 
more closely resembles the spectra of SNe~2011fe and 2011iv. Meanwhile, 
ASASSN-14lp and SN~2021fxy have nearly identical spectra between 3000-3600~\AA, 
with SN~2017erp showing lower NUV fluxes than any other SN~Ia with 
significant MUV suppression. Returning to \autoref{fig:uv_colour_evo}, 
all four SNe are members of the NUV-red group, following the same 
general evolutionary track, albeit with significant scatter. ASASSN-14lp 
and SN~2017erp generally appear redder than SNe~2013dy and 2021fxy at 
all epochs. 

\begin{figure*}
  \centering
  \includegraphics[width=\textwidth]{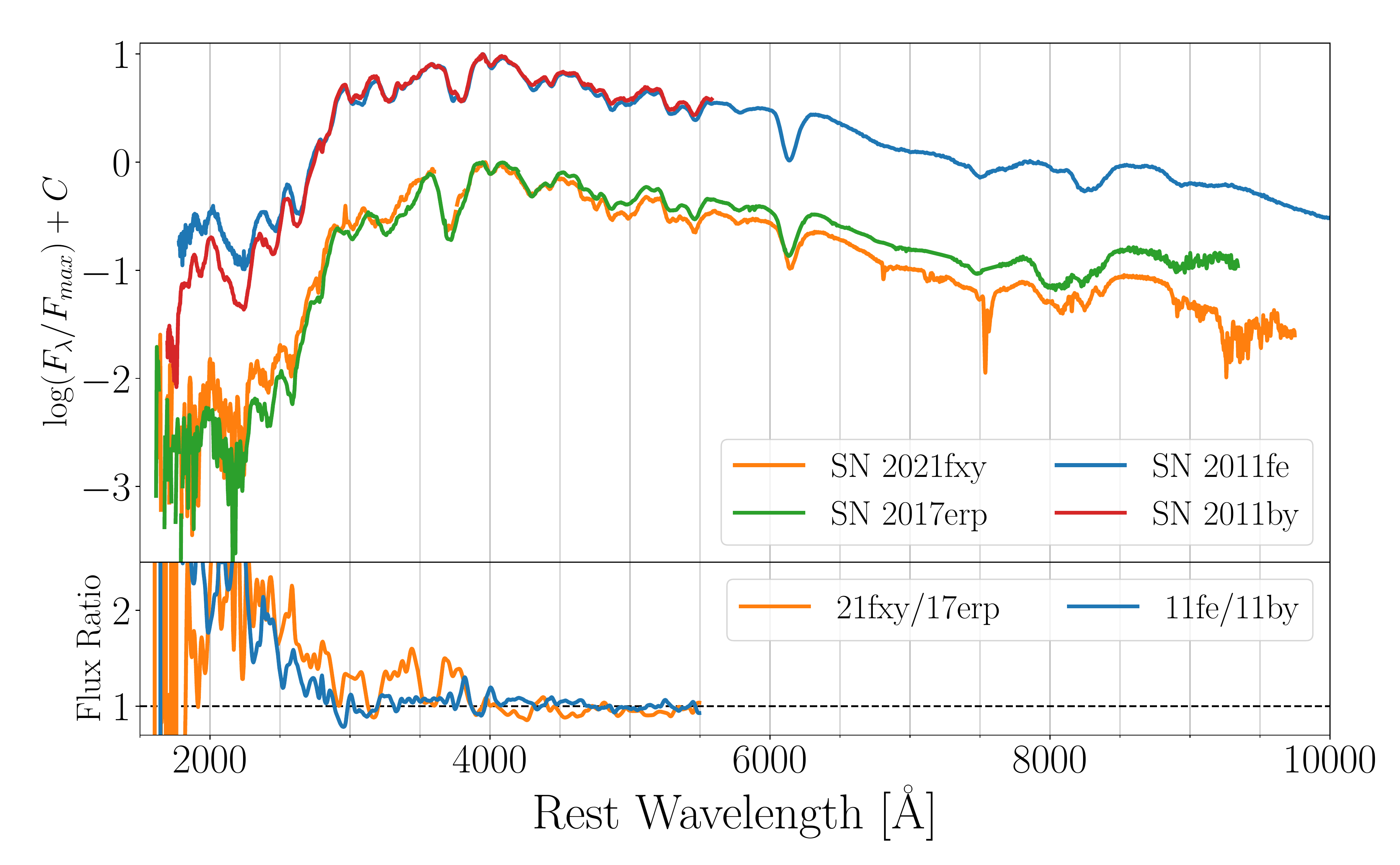}
  \caption{\textit{Top Panel}: Combined UV/Optical spectra of SN~2021fxy and
    2017erp at maximum light compared to those of SN~2011fe and 2011by.
    \textit{Bottom Panel}: Flux ratio of SN~2021fxy/2017erp compared to that of
    SN~2011fe/2011by, as determined from the {\it HST} portions of the spectra.}
  \label{fig:21fxy_17erp_uvo_comp}
\end{figure*}

The bigger picture becomes even less clear when we begin to consider 
the optical spectral properties of the MUV-suppressed SNe. SNe~2013dy, 
ASASSN-14lp, and 2021fxy are all members of the SS class, while 
SN~2017erp belongs to the CN class. While the spectral behavior of 
SNe~2013dy and ASASSN-14lp are typical of members of the SS subgroup, 
SNe~2017erp and 2021fxy share many characteristics, including nearly 
identical optical spectra and light curve parameters, 
($s_{BV,21fxy} = 0.99 \pm 0.03$, $\Delta m_{15}(B)_{21fxy} = 1.05 \pm 0.06$; 
$s_{BV,17erp} = 0.99 \pm 0.03$, $\Delta m_{15}(B)_{17erp} = 1.05 \pm 0.06$).
When we expand this comparison to include spectra both before and after 
maximum light, we find that both object's spectra show similar feature 
velocities, line profiles, and line strengths for nearly all of their 
lines throughout their evolution. The notable exceptions to this being 
the \ion{Si}{ii} $\lambda5972$ and $\lambda6355$ lines. While it is 
tempting to establish a familial relationship between 2021fxy and 2017erp, 
given the numerous similarities is their spectral and photometric properties,
unlike the ``twin'' supernovae 2011by and 2011fe, SNe~2017erp and 2021fxy 
are not members of the same Branch group (see again \autoref{fig:branch_diagram}, 
also \autoref{sec:synow}); nor are they ``siblings" hosted in the same galaxy.
SNe~2021fxy and 2017erp also show significantly different continuum levels 
in the optical, as seen in \autoref{fig:21fxy_17erp_uvo_comp}, which persists 
throughout the photospheric phase. Despite these differences, from the 
{\it HST} spectra, SNe~2021fxy and 2017erp produce a flux ratio bluewards of 
$\sim5600$~\AA~that is comparable to the one between SNe~2011by and 2011fe 
at maximum light, as seen in the lower panel of \autoref{fig:21fxy_17erp_uvo_comp}. 
As such, we analyse the {\it HST} spectra of SNe~2017erp and 2021fxy in an 
attempt to determine the source of their UV flux differences.

\subsection{Comparing SNe~2017erp and 2021fxy}

\subsubsection{Optical Spectra Similarities}

Early on, both supernovae show similar high-velocity (HV) \ion{Si}{ii}
lines with the velocity of the feature in SN~2017erp declining from
$-21,600$~km~s$^{-1}$ to $-18,000$~km~s$^{-1}$ from the first spectrum
taken at $-17.0$~days to the one taken at $-14.1$~days. The first spectrum
of SN~2021fxy at $-13.9$~days shows  HV \ion{Si}{ii} at $-18,200$~km~s$^{-1}$. 
Roughly a week later, the HV \ion{Si}{ii} has mostly faded 
from both supernovae. However, the $-5.9$~day spectrum of SN~2021fxy shows 
a flat-topped emission peak associated with a feature detached from the
underlying photosphere; compared to the traditional P-Cygni profile seen 
in the $-8.5$~day spectrum of SN~2017erp, where the feature begins forming
at the photosphere (see \autoref{sec:synow} for further details). 
In SN~2021fxy, the \ion{Si}{ii} remains
detached throughout the photospheric phase, until it begins to be
polluted by Fe lines, which first appear around $+11$/$+12$~days, and are 
clearly present at $+15.7$~days. In contrast, the \ion{Si}{ii} feature in 
SN~2017erp does not detach from the photosphere until $+6$~days, and does
not begin to show signs of the photosphere entering the Fe-rich inner
region until $+17$~days. Taken together, these differences support the
classification of SN~2021fxy as a SS
object. \citet{Nugent:Phillips:1995} showed that the 
Branch sequence (neglecting BL) is driven by differences in
temperature, with CL being coolest and SS being hottest. To zeroth
order these temperature differences may be associated with the total
amount of \nnni produced in the explosion. Thus, in a near Chandrasekhar
mass progenitor scenario  we expect SS to produce somewhat more \nnni
and somewhat less silicon, leading to  more rapid evolution in the
\ion{Si}{ii} features for SS as compared with CN supernovae.

\begin{figure}
  \centering
  \includegraphics[width=\columnwidth]{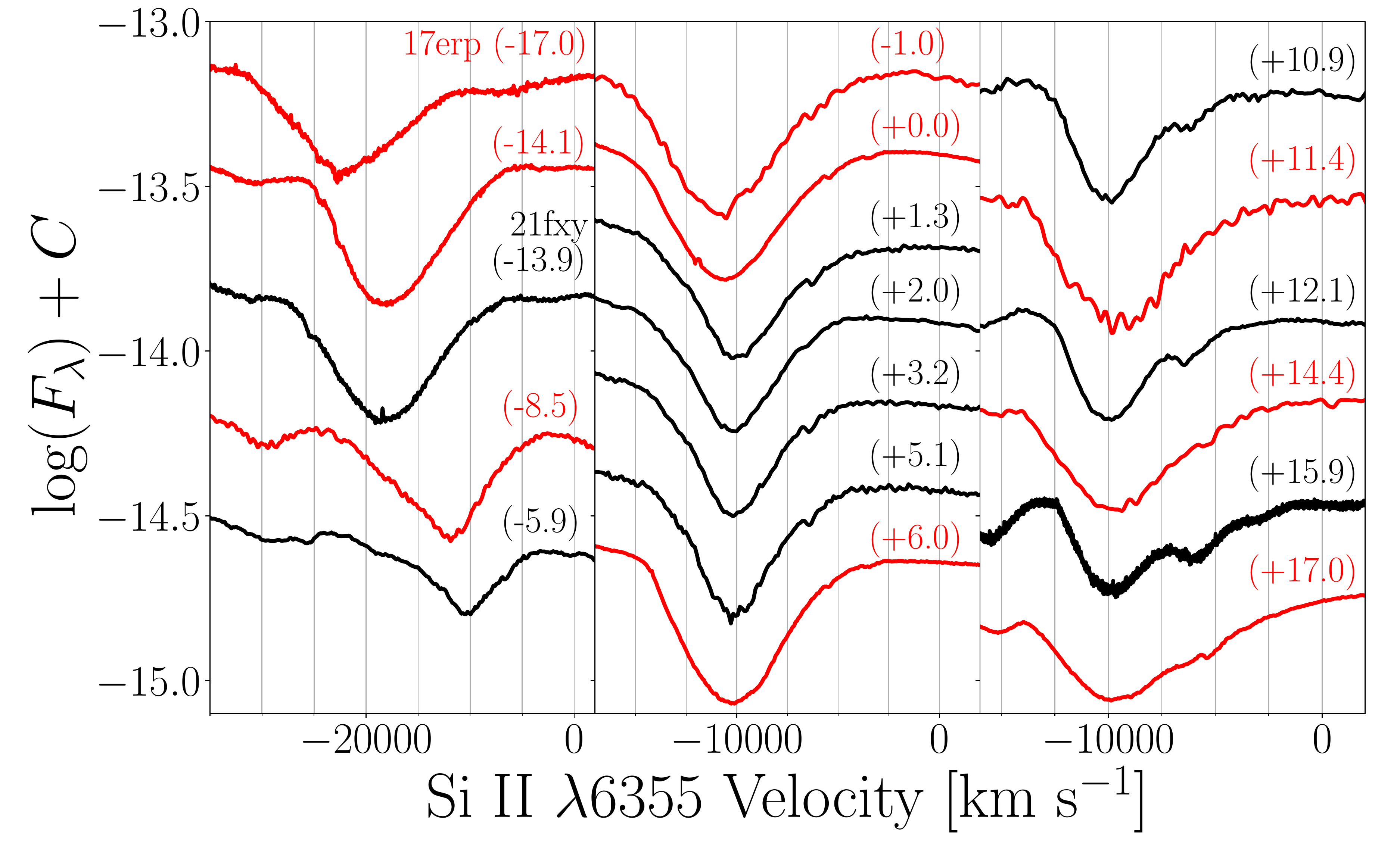}
  \caption{Comparison of \ion{Si}{ii} $\lambda6355$ line evolution in
  SN~2021fxy and SN~2017erp.}
  \label{fig:21fxy_17erp_si2}
\end{figure}

\subsubsection{UV Flux Differences}

Due to the complicated relationships between physical factors 
(metallicity, density structure, and luminosity) that have strong 
impacts on UV spectral formation, attributing UV spectral variations
to an individual factor is difficult. Therefore, we examine the flux 
ratios from self-consistent sets of models, varying one factor at a 
time, to place limits on the relative differences in metallicity and
luminosity between SNe~2017erp and 2021fxy by assuming the observed 
UV differences are caused solely by an individual factor.

\begin{figure*}
  \centering
  \includegraphics[width=\textwidth]{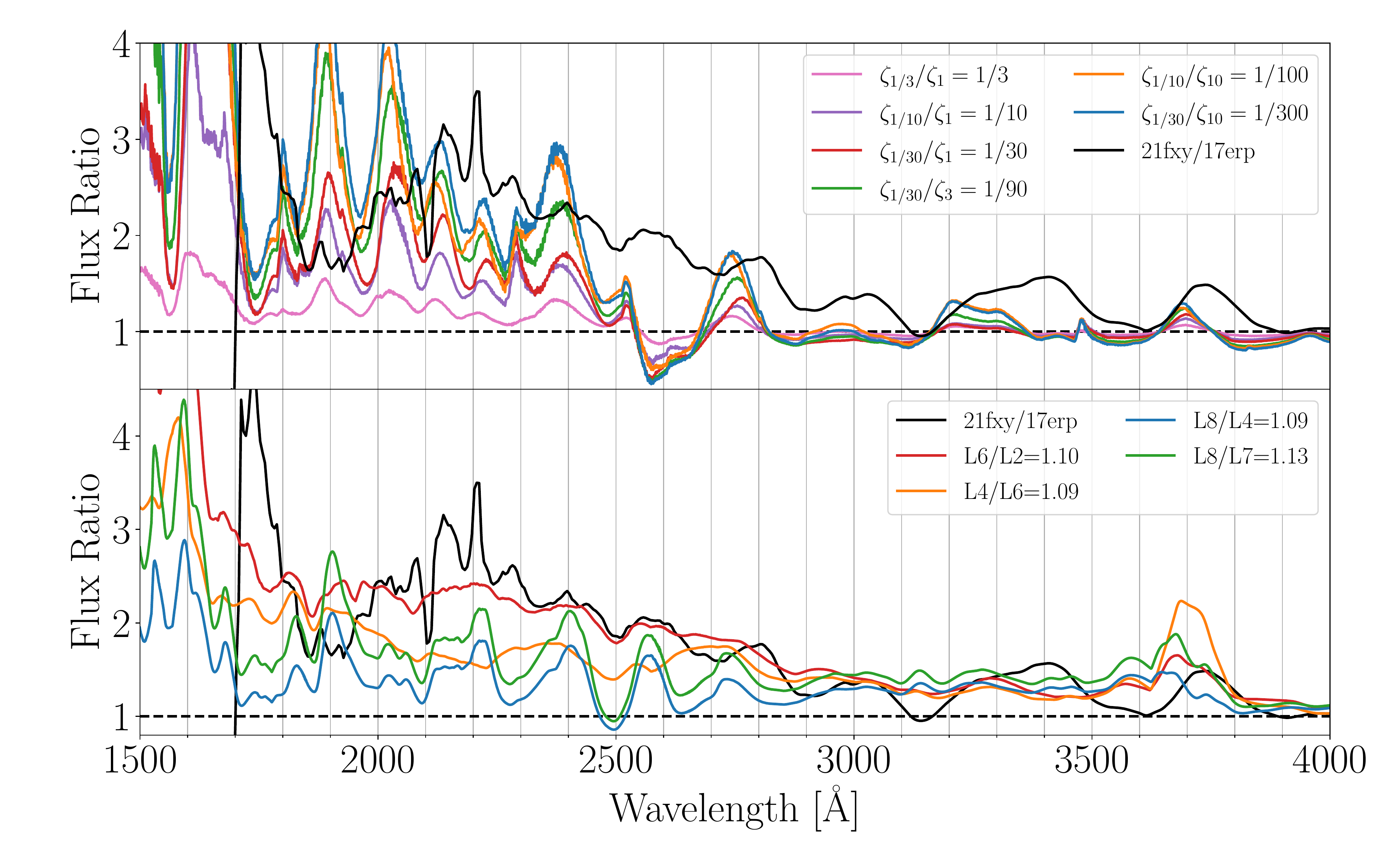}
  \caption{\textit{Top Panel}: Flux ratios of the Day 15 spectral models 
    from \citet{Lentz2000} with the flux ratio of SN~2021fxy to SN~2017erp 
    over-plotted. \textit{Bottom Panel}: Flux ratios for selected models 
    of \citet{DerKacy2020} compared to the flux ratio of SN~2021fxy to 
    SN~2017erp.}
  \label{fig:muv_models}
\end{figure*}

\paragraph{Metallicity Variations}
Using the $t=15$ day models of \cite{Lentz2000}, we can explore 
differences in the relative progenitor metallicity as the source of the 
UV differences between the two supernovae. These models are NLTE
simulations based on W7, with the abundances of all elements heavier 
than oxygen in the unburned C+O layer ($v \gtrsim 15000$ km s$^{-1}$) 
scaled by factors of $\zeta = $ 1/30, 1/10, 1/3, 3, and 10 relative to 
solar metallicity. As previously noted by \citet{Foley2013}, the flux 
ratios from models with the same metallicity ratio produce the same general 
trend across the UV, and therefore are only able to infer relative 
metallicities between the two SNe, not differentiate the absolute 
metallicities of the supernovae. In exploring the impact of relative 
metallicity differences between SNe~2021fxy and 2017erp, we examine only 
the region between 2000-2500~\AA, as spectral formation in this region is 
almost entirely determined by iron group elements 
\citep[see Figure 8 of][]{DerKacy2020}. No combination of any two Lentz models 
is able to reproduce the flux ratio of 2021fxy/2017erp across the entire 
wavelength range, in part due to the diminishing ability of these models 
to distinguish between increasingly large differences in the relative
metallicity. The best match to the 2021fxy/2017erp flux ratio is produced 
by the $\zeta_{1/30}/\zeta_{10}$ curve ($\chi^{2} = 166.87$, $\chi^{2}_\nu = 1.7$). 
If we instead fit over the entire {\it HST} spectra with $\lambda > 1800$~\AA{}, 
we find that we can no longer distinguish between the flux ratio curves produced 
by the $1/300$ ($\chi^{2} = 328.19$), $1/100$ ($\chi^{2} = 328.23$), and 
$1/90$ ($\chi^{2} = 341.10$) metallicity ratios. This result matches what we 
see in the top panel of \autoref{fig:muv_models}, as each of these three curves 
are virtually indistinguishable redward of 2500~\AA{}, and only distinguished
by small variations in the height of a few peaks between $2000-2500$~\AA{} where 
metallicity differences should be most apparent. 

\paragraph{Luminosity Variations}
The bottom panel of \autoref{fig:muv_models} shows selected flux ratios of 
models from \cite{DerKacy2020} compared to the flux ratio of 2021fxy/2017erp.
Analysis of these models reveal that unlike the flux ratios generated from
the models of \cite{Lentz2000}, the flux ratios generated from the
\cite{DerKacy2020} models are sensitive to both the relative
luminosity difference and absolute luminosity of the model. This is
due to the luminosity differences inducing temperature variations in
the outer ejecta that alter the shape of the underlying continuum, as
well as the excitation and ionization states of the outer ejecta. The
differences produce different strengths and locations of spectral
features in the ejecta, resulting in unique flux ratio curves
dependent on the luminosities of the two spectra in the ratio. In
effect this is an application of the Spectral-fitting Expanding
Atmosphere Method \citep[SEAM;][]{Mitchell:2002,Baron:2004,Dessart:2010}.
We find good agreement between the flux ratios produced from the L6/L2
models ($\chi^{2} = 190.77$, $\chi^{2}_\nu = 0.25$) and the flux ratio 
of 2021fxy/2017erp, implying that SN~2021fxy has a peak bolometric luminosity 
10\% higher than that of SN~2017erp.  

However, we caution against too strong an interpretation of these results 
as the model and input luminosities used in \citet{DerKacy2020} were only
simulated at one epoch and were chosen because they best reproduce SN~2011fe,
which is known to have the bluest UV minus optical colours among SNe~Ia with 
UV spectra \citetext{\citealp{Brown2017}; P.~Brown, et al., in preparation}. 
Furthermore, both the best-fitting luminosity and metallicity ratio curves 
only broadly capture the observed differences between SNe~2017erp and 2021fxy 
in the mid-UV, where the differences in both properties are expected to 
be the greatest. It is only when the differences in flux from the remainder 
of the UV and optical are examined (see again 
\autoref{fig:21fxy_17erp_uvo_comp}), we find that the behavior is more 
consistent with those expected from differences in luminosity than 
metallicity \citep{Lentz2000,DerKacy2020}. In the case of SNe~2011by
and 2011fe, \citet{Graham2015} found that the inclusion of additional
information, such as UV time series spectra and nebular spectra, may 
disfavor certain explanations like metallicity differences as the root 
cause of UV variations; despite the inability of any single parameter to 
adequately reproduce the observed UV differences without inducing unobserved
differences in the optical. \citet{Foley2020} also showed that there are
significant differences in the luminosities of SNe~2011by and 2011fe after
re-calibrating the distance to NGC~3972 (host of SN~2011by) which may be 
explained by multiple mechanisms. Additional work to verify the relationship 
between UV spectra and SNe~Ia properties like metallicity and luminosity 
across more models and epochs in a self-consistent manner is currently 
underway.

\subsection{Sibling's Analysis}

Detailed analyses of SNe~Ia siblings (e.g. two or more SNe~Ia hosted in the
same galaxy) allow us to test many of our assumptions about SNe~Ia as cosmological 
distance indicators. By virtue of sharing the same host, many of the factors that 
increase the scatter in cosmological distance measurements are eliminated, including 
dependencies on properties of the host galaxy such as host mass and metallicity,
and peculiar velocities \citep{Sullivan2010,Brown2014,Burns2020,Scolnic2020,Hoogendam2022}.

The one confirmed sibling to SN~2021fxy, SN~2002dj, was studied in depth by 
\cite{Pignata2008}, and determined to be a Ia-BL within the Branch scheme with 
several similarities to SN~2002bo. Using \texttt{SNooPy}, we fit the photometry of SN~2002dj, 
with the results compared to those of SN~2021fxy in \autoref{tab:sib_lc_params}. 
From the results, we find that the implied distance modulus to NGC 5018 agree to
$1.2\sigma$,  within the average $\Delta\mu$ of other sibling SNe~Ia studied by 
\citet{Burns2020}. Both values are also consistent with the redshift derived value to less
than $1\sigma$. The inferred host extinctions are consistent at the $1.3\sigma$ level.
The different estimates of $E(B-V)$ are likely due to different local environments in 
the vicinity of the two SNe. SN~2002dj was found to be coincident with an extended 
emission region appearing as a warped disc covering portions of NGC 5018, and is 
associated with regions of star formation \citep{Pignata2008,Goudfrooij1994}. 
SN~2021fxy exploded in a part of NGC 5018 unassociated with this emission region.

\begin{table}
\centering
\caption{Comparison of Light Curve Parameters\label{tab:sib_lc_params}}
\begin{tabular}{ccc}
  \hline
  Parameter & SN~2021fxy & SN~2002dj \\
  \hline
  $t_{max}$ (MJD) & $59305.1 \pm 0.3$ & $52450.9 \pm 0.4$ \\
  $s_{BV}$ & $0.99 \pm 0.04$ & $0.94 \pm 0.04$ \\
  $E(B-V)_{host}$ (mag)& $0.02 \pm 0.06$ & $0.10 \pm 0.06$ \\
  $\mu$ (mag) & $32.87 \pm 0.09$ & $32.97 \pm 0.09$ \\
  \hline
\end{tabular}
\end{table}

\section{Conclusions}\label{sec:conclusion}

We present detailed photometric and spectroscopic followup of SN~2021fxy, 
a SN~Ia discovered in NGC~5018 for which we also obtained UV photometry with
{\it Swift/UVOT} and UV spectroscopy with {\it HST}/STIS. Ground-based 
photometric and spectroscopic observations were obtained as part of the Precision 
Observations of Infant Supernova Explosions 
(\href{https://poise.obs.carnegiescience.edu/}{POISE}, \citealp{Burns2021}), 
the Aarhus-Barcelona FLOWS, and NUTS2 collaborations. Our observations reveal 
that SN~2021fxy has a normal light curve consistent with a normal-bright 
SN~Ia. The $B-V$ colours of SN~2021fxy are red in the first days after explosion 
but rapidly evolve blueward. Optical spectra show SN~2021fxy to be a member of 
the SS group, located near the border of the CN and SS groups within the Branch 
diagram. In contrast to other SS objects from the \citet{Stritzinger2018} sample, 
SN~2021fxy is the only known SS object to show an early-time red $B-V$ colour, 
suggesting that early colour diversity a complex phenomenon not captured in SN~Ia 
sub-classification schemes. A better understanding of the diversity of early-time 
colours and their connections to observational properties, progenitor systems, and 
explosion mechanisms are a key scientific goal of the POISE collaboration. 

UV spectra show that when compared to other spectroscopically normal SNe~Ia, 
SN~2021fxy is a member of a group of objects with flux suppression in the 
mid-UV, which cannot be explained by host reddening alone. Objects with MUV 
flux suppression all belong to the NUV-red group of SNe~Ia, possess MUV 
features that are bluer than their non-suppressed counterparts and HV 
components in their \ion{Ca}{ii} H\&K lines that are dominant over the 
photospheric components, as measured by the quantity $R_{HVF}$. One potential 
cause of this suppression is an increased effective opacity in the mid-UV 
from IGEs at higher velocities, which would imply a continuous distribution 
of MUV flux values in SNe~Ia. However, the presence of the HV Ca features 
could indicate that shells of material within the ejecta are responsible for 
the additional line blanketing. In either case, more UV spectral observations 
of SNe~Ia are needed, especially of BL or High-velocity supernovae, in order
to determine the physical mechanism responsible for the MUV flux suppression. 

Among those SNe~Ia with MUV flux suppression, SNe~2021fxy and 2017erp show 
remarkable similarities despite belonging to the SS and CN Branch subgroups
respectively, which allow us to probe the mechanisms responsible for 
variations between different MUV-suppressed objects. We find that the flux 
differences between SNe~2021fxy and 2017erp in the UV are comparable in 
size to those between SNe~2011by and 2011fe, but are instead likely due 
to variations in the intrinsic luminosity differences between the two SNe; 
not metallicity differences as has been suggested for SNe~2011by and 2011fe. 
Further modeling to better understand the impact of different physical 
mechanisms which contribute to UV spectral formation, and which observational 
quantities best measure this diversity are ongoing.

\section*{Acknowledgements}

E.B. and J.D. are supported in part by NASA grant 80NSSC20K0538. Some 
of the calculations presented here were performed at the
H\"ochstleistungs Rechenzentrum Nord (HLRN), at the National Energy 
Research Supercomputer Center (NERSC), which is supported by the 
Office of Science of the U.S.  Department of Energy under Contract 
No. DE-AC03-76SF00098 and at the OU Supercomputing Center for Education \& 
Research (OSCER) at the University of Oklahoma (OU). We thank all these 
institutions for a generous allocation of computer time.M.D.S. and E.K. 
are supported by an Experiment grant (\# 28021) from the Villum FONDEN, 
and by a project 1 grant (\#8021-00170B) from the Independent Research 
Fund Denmark (IRFD).L.G. acknowledges financial support from the Spanish 
Ministerio de Ciencia e Innovaci\'on (MCIN), the Agencia Estatal de 
Investigaci\'on (AEI) 10.13039/501100011033, and the European Social Fund 
(ESF) "Investing in your future" under the 2019 Ram\'on y Cajal program 
RYC2019-027683-I and the PID2020-115253GA-I00 HOSTFLOWS project, from 
Centro Superior de Investigaciones Cient\'ificas (CSIC) under the PIE 
project 20215AT016, and the program Unidad de Excelencia Mar\'ia de 
Maeztu CEX2020-001058-M. BJS is also supported by NSF grants AST-1907570, 
AST-1920392 and AST-1911074. AR acknowledges support from ANID BECAS/DOCTORADO 
NACIONAL 21202412.
Based on observations with the NASA/ESA Hubble Space Telescope
obtained at the Space Telescope Science Institute, which is operated 
by the Association of Universities for Research in Astronomy, 
Incorporated, under NASA contract NAS5-26555. Support for Program number 
16221 was provided through a grant from the STScI under NASA contract 
NAS5-26555. This paper includes data gathered with the 6.5 meter Magellan 
Telescopes located at Las Campanas Observatory, Chile. We thank the students 
of the Spring 2021 Advanced Observatory Methods class at OU, Mukremin Kilic, 
and the staff at Apache Point Observatory including Russet McMillan and Candace 
Gray for obtaining spectra of SN~2021fxy during their telescope training on 
April 1 and 3, 2021. Based on observations made with the Nordic Optical Telescope 
(NOT), operated by the Nordic Optical Telescope Scientific Association at the 
Observatorio del Roque de los Muchachos, La Palma, Spain, of the Instituto 
de Astrofisica de Canaries.  Data obtained with the NOT by NUTS2 was done in 
part due to funds from the Instrument-center for Danish Astrophysics (IDA).

\section*{Data Availability}

Some data underlying this article are available in the article, with other
data available in its online supplementary material. Full photometry can be 
found in \autoref{sec:phot}. The spectra presented in this paper are available 
from the 
\href{https://poise.obs.carnegiescience.edu/data}{data downloads page of the POISE site}. 
They have also been archived at \href{https://www.wiserep.org/}{WISeREP}.



\bibliographystyle{mnras}
\bibliography{ms} 




\appendix

\section{Photometric Data}\label{sec:phot}

\begin{table*}
\centering
\caption{Log of {\it Swift} Photometry for SN~2021fxy \label{tab:swift_phot}}
\begin{tabular}{ccccccc}
  \hline
  MJD & UVM2 [mag] & UVW1 [mag] & UVW2 [mag] & U [mag] & B [mag] &V [mag] \\
  \hline
  59291.39 & --- & --- & --- & --- & --- & 16.53(0.12) \\
  59291.45 & --- & 18.74(0.11) & --- & 17.24(0.007) & 16.70(0.06) & --- \\
  59292.13 & --- & --- & --- & --- & --- & 15.85(0.08) \\
  59292.18 & --- & 18.74(0.11) & --- & 16.89(0.07) & --- & --- \\
  59292.19 & --- & --- & --- & --- & 16.12(0.06) & --- \\
  59293.37 & --- & 17.33(0.13) & --- & 15.89(0.09) & 15.59(0.07) & 15.52(0.10) \\
  59300.12 & --- & 15.48(0.06) & 17.07(0.09) & 13.80(0.04) & 14.08(0.04) & 14.07(0.06) \\
  59300.24 & --- & --- & 17.14(0.11) & --- & --- & --- \\
  59300.32 & --- & --- & 17.15(0.07) & --- & --- & --- \\
  59300.35 & 18.82(0.27) & --- & --- & --- & --- & --- \\
  59304.59 & --- & 15.40(0.07) & 16.81(0.10) & 13.73(0.04) & 13.91(0.04) & 13.78(0.05) \\
  59309.77 & --- & 15.77(0.08) & --- & 14.06(0.05) & --- & --- \\
  59309.78 & --- & --- & 17.22(0.20) & --- & 14.03(0.04) & --- \\
  \hline
\end{tabular}
\end{table*}

\begin{table*}
\centering
\caption{Log of Optical Photometry of SN~2021fxy}
\begin{tabular}{cccccccc}
  \hline
  MJD & $u$ [mag] & $B$ [mag] & $V$ [mag] & $g$ [mag] & $r$ [mag] & $i$ [mag] & Telescope \\
  \hline
  59291.12 & 18.54(0.04) & 17.18(0.01) & 16.65(0.01) & 16.93(0.01) & 16.66(0.01) & 17.06(0.02) & Swope \\
  59291.32 & 18.38(0.04) & 16.97(0.01) & 16.54(0.01) & 16.79(0.01) & 16.55(0.01) & 16.93(0.02) & Swope \\
  59294.10 & 16.18(0.03) & 15.27(0.01) & 15.16(0.01) & 15.21(0.01) & 15.20(0.01) & 15.53(0.01) & Swope \\
  59294.31 & 16.07(0.03) & 15.19(0.01) & 15.10(0.01) & 15.13(0.01) & 15.13(0.01) & 15.45(0.01) & Swope \\
  59295.11 & 15.69(0.03) & 14.93(0.01) & 14.87(0.01) & 14.88(0.01) & 14.90(0.01) & 15.21(0.01) & Swope \\
  59295.34 & 15.61(0.03) & 14.81(0.01) & 14.80(0.01) & 14.78(0.01) & 14.80(0.01) & 15.11(0.01) & Swope \\
  59295.40 & 15.57(0.03) & 14.85(0.01) & 14.78(0.01) & 14.79(0.01) & 14.83(0.01) & 15.13(0.01) & Swope \\
  59296.09 & 15.32(0.03) & 14.66(0.01) & 14.62(0.01) & 14.62(0.01) & 14.66(0.01) & 14.94(0.01) & Swope \\
  59296.27 & 15.25(0.03) & 14.54(0.01) & 14.57(0.01) & 14.55(0.01) & 14.61(0.01) & 14.91(0.01) & Swope \\
  59296.39 & 15.25(0.03) & 14.56(0.01) & 14.57(0.01) & 14.54(0.01) & 14.58(0.01) & 14.87(0.02) & Swope \\
  59297.25 & 15.02(0.03) & 14.42(0.01) & 14.41(0.01) & 14.40(0.01) & 14.45(0.01) & 14.75(0.01) & Swope \\
  59298.25 & 14.85(0.03) & 14.27(0.01) & 14.27(0.01) & 14.24(0.01) & 14.30(0.01) & 14.65(0.01) & Swope \\
  59299.11 & --- & 14.09(0.01) & 14.21(0.01) & 14.15(0.01) & 14.23(0.01) & --- & LCOGT \\
  59299.25 & 14.72(0.03) & 14.17(0.01) & 14.16(0.01) & 14.13(0.01) & 14.19(0.01) & 14.56(0.01) & Swope \\
  59300.20 & 14.65(0.03) & 14.11(0.01) & 14.08(0.01) & 14.06(0.01) & 14.13(0.01) & 14.56(0.01) & Swope \\
  59300.93 & --- & 13.89(0.01) & 14.03(0.01) & 13.97(0.01) & 14.08(0.01) & --- & LCOGT \\
  59301.19 & 14.59(0.03) & 13.96(0.02) & 14.01(0.01) & --- & --- & 14.52(0.01) & Swope \\
  59302.24 & 14.55(0.03) & 13.96(0.01) & 13.94(0.01) & 13.94(0.01) & 14.02(0.01) & 14.56(0.01) & Swope \\
  59303.19 & 14.54(0.03) & 13.93(0.01) & 13.92(0.01) & 13.90(0.01) & 13.97(0.01) & 14.55(0.02) & Swope \\
  59304.23 & 14.58(0.03) & 13.93(0.01) & 13.90(0.01) & 13.90(0.01) & 13.95(0.01) & 14.60(0.01) & Swope \\
  59305.25 & 14.57(0.03) & 13.95(0.01) & 13.90(0.01) & 13.89(0.01) & 13.94(0.01) & 14.62(0.01) & Swope \\
  59306.16 & 14.63(0.03) & 13.93(0.01) & 13.89(0.01) & 13.88(0.01) & 13.91(0.01) & 14.63(0.01) & Swope \\
  59307.14 & 14.68(0.03) & 13.97(0.01) & 13.91(0.01) & 13.90(0.01) & 13.94(0.01) & 14.67(0.01) & Swope \\
  59309.26 & 14.81(0.03) & 14.03(0.01) & 13.95(0.01) & 13.96(0.01) & 13.96(0.01) & 14.72(0.01) & Swope \\
  59310.21 & 14.89(0.03) & 14.08(0.01) & 13.98(0.01) & 14.00(0.01) & 14.01(0.01) & 14.78(0.01) & Swope \\
  59311.23 & 14.96(0.03) & 14.04(0.01) & 13.96(0.01) & 14.01(0.01) & 14.01(0.01) & 14.77(0.01) & Swope \\
  59312.30 & 15.10(0.05) & 14.21(0.02) & 14.04(0.01) & 14.09(0.01) & 14.10(0.01) & 14.85(0.01) & Swope \\
  59312.30 & 15.14(0.08) & 14.21(0.02) & 14.04(0.01) & 14.09(0.01) & 14.10(0.01) & 14.85(0.01) & Swope \\
  59313.23 & 15.14(0.03) & 14.20(0.01) & 14.06(0.01) & 14.11(0.01) & 14.12(0.01) & 14.91(0.01) & Swope \\
  59313.46 & --- & 14.24(0.01) & 14.13(0.01) & 14.20(0.01) & 14.21(0.01) & --- & LCOGT \\
  59314.23 & 15.26(0.03) & 14.36(0.01) & 14.13(0.01) & 14.21(0.01) & 14.23(0.01) & 15.01(0.01) & Swope \\
  59315.06 & 15.32(0.03) & 14.41(0.01) & 14.18(0.01) & 14.26(0.01) & 14.30(0.01) & 15.08(0.01) & Swope \\
  59316.29 & 15.50(0.03) & 14.49(0.01) & 14.24(0.01) & 14.32(0.01) & 14.36(0.01) & 15.13(0.01) & Swope \\
  59317.23 & 15.62(0.03) & 14.52(0.01) & 14.28(0.01) & 14.37(0.01) & 14.42(0.01) & 15.22(0.01) & Swope \\ 
  59324.29 & --- & --- & 14.74(0.01) & 15.07(0.01) & 14.76(0.01) & 15.23(0.01) & LCOGT \\
  59325.85 & --- & 15.53(0.01) & 14.82(0.01) & 15.18(0.01) & 14.75(0.01) & 15.14(0.01) & LCOGT \\
  59337.42 & --- & 16.55(0.02) & 15.59(0.01) & 16.21(0.01) & 15.28(0.01) & 15.31(0.01) & LCOGT \\
  59342.27 & --- & 16.84(0.03) & 15.84(0.02) & 16.48(0.01) & 15.57(0.01) & 15.67(0.02) & LCOGT \\
  59345.72 & --- & 16.89(0.02) & 16.00(0.01) & 16.55(0.01) & 15.75(0.01) & 15.84(0.01) & LCOGT \\
  59350.71 & --- & 17.05(0.02) & 16.10(0.01) & 16.67(0.01) & 15.93(0.01) & 16.07(0.01) & LCOGT \\
  59352.79 & --- & 17.08(0.01) & 16.18(0.01) & 16.75(0.01) & 15.99(0.01) & 16.16(0.01) & LCOGT \\
  59361.82 & --- & 17.12(0.10) & 16.37(0.05) & 16.92(0.03) & 16.29(0.01) & 16.35(0.09) & LCOGT \\
  59364.74 & --- & 17.29(0.02) & 16.48(0.01) & 16.93(0.01) & 16.38(0.01) & 16.61(0.02) & LCOGT \\
  59365.58 & --- & 17.31(0.02) & 16.53(0.01) & 16.97(0.01) & 16.43(0.01) & 16.67(0.01) & LCOGT \\
  59374.89 & --- & 17.39(0.02) & 16.76(0.02) & 17.10(0.01) & 16.69(0.01) & 16.94(0.02) & LCOGT \\  
  \hline
\end{tabular}
\end{table*}

\clearpage

\section{\texttt{SYNOW} Fits of SN~2021fxy} \label{sec:synow}

\texttt{SYNOW} is a highly parameterized code designed to simulate supernova
spectra, to assist in the identification of spectral lines and estimation of 
the both photospheric velocity and velocity interval of ions within the supernova
ejecta. It makes simple assumptions about the supernova, including spherical
symmetry, homologous expansion, line formation via resonance scattering in the
Sobolev approximation, and a sharp photosphere emitting a blackbody continuum
to calculate a synthetic spectra. Key user defined parameters include the 
temperature of the blackbody continuum, photospheric velocity, and the reference
line optical depth, e-folding velocity, velocity extent, and the Boltzmann 
excitation temperature for each ion included in the fit. The best--fitting spectrum
is then determined via ``chi-by-eye", as is the community standard. More 
information on \texttt{SYNOW} can be found in \citet{Jeffery1990} and 
\citet{Branch2005,Branch2006}.

As stated above, SNe~2021fxy and 2017erp show numerous similarities. Both have
suppressed flux in the mid-UV and show features that are nearly identical in velocity,
line profile, and line depth. However, SN~2021fxy evolves through its photospheric 
phase faster than 2017erp, as measured by the \ion{Si}{ii} $\lambda 6355$ line, and the
two SNe are members of different Branch groups. Using \texttt{SYNOW}, we can investigate 
just how similar the ejecta of the two SNe are. 

Our generalized fitting procedure is as follows. After assuming a blackbody temperature
$T_{bb}$, we fit the \ion{Si}{ii} features, assuming that the photospheric velocity 
($v_{phot}$) is the same as the $v_{min}$ of \ion{Si}{ii}. With the photospheric velocity (PV) 
established, we fit ions of other intermediate mass elements (IMEs), including \ion{Ca}{ii}, 
\ion{S}{ii}, \ion{Mg}{ii}, etc., including any high velocity (HV) components. Once initial
fits of the IMEs are complete, we add the important ions arising from the iron group 
elements (IGEs), including \ion{Fe}{ii}, \ion{Fe}{iii}, \ion{Co}{ii}, and \ion{Ni}{ii}, 
revising our IME parameters as necessary to fit blended features. We assume an
excitation temperature of ($T_{exec} = 10000$ K) unless stated otherwise. The full set 
of input parameters are listed in \Autoref{tab:synow_17erp,tab:synow_21fxy}. We briefly 
summarize our important findings below.

As previously shown in \autoref{fig:21fxy_17erp_si2}, several epochs of SN~2021fxy 
show broad, flat-topped emission profiles characteristic of line formation occurring
in a region detached from the photosphere \citep{Jeffery1990}. However, because of our 
assumption that $v_{phot} = v_{min,Si}$, we only find one epoch of SN~2021fxy where
\ion{Si}{ii} is clearly detached. Most likely, the \ion{Si}{ii} in the preceding epochs
is detached by $\lesssim 1000$ km s$^{-1}$, as this represents the $3\sigma$ error in
our velocity measurements. This is supported by the appearance of a weak \ion{C}{ii} 
$\lambda 6580$ line at $+5.1$ days in SN~2021fxy, which serves to further flatten
the emission peak. We find further support for this idea by examining the 
\ion{Si}{iii} lines, which is also detached from the photosphere at $+5.1$ days. As none of 
the Si lines in the SN~2017erp fits are detached, our \texttt{SYNOW} fits support 
our finding that SN~2021fxy evolves through its photospheric phase faster than 
SN~2017erp.

The \ion{Ca}{ii} lines proved particularly difficult to fit well. In addition to 
many of the NIR triplet features showing flat-topped emission peaks similar to
the \ion{Si}{ii} lines, both the H\&K lines and the NIR triplets often required 
multiple detached or HV components to accurately represent the feature. We were 
able to distinguish the different HV components through their different $v_{max}$
values. These narrow regions may be indications of a series of shells in the outer
layers of the SNe ejecta. Yet, the numerous components often resulted in fits that 
were not able to reproduce both features accurately. In SN~2021fxy, we were able to
obtain good fits to both the \ion{Ca}{ii} features in all epochs except $+1.3$d and 
$+10.9$d, where the NIR triplet is preferentially fit. For SN~2017erp, the NIR triplet 
is preferentially fit in the $-17.0$d and $-14.1$d spectra due to incomplete coverage
of the H\&K features, and is preferentially fit in the $-8.5$d and $-1.0$d spectra.

\ion{S}{ii} lines are present from the earliest epochs in both SNe. The 
features grow stronger in both SNe, peaking in strength near maximum light 
before weakening significantly by $\sim +11$ days. The strength of \ion{S}{ii} 
is correlated with the photospheric temperature, however the response is
both non-monotonic and strongly influenced by non-local thermodynamic
equilibrium (NLTE) effects
\citep{Nugent:Phillips:1995}. Therefore, although the excitation temperatures 
of the \ion{S}{ii} lines increase in both SNe, also peaking near maximum light 
it is difficult to discern whether this accurately captures the physics. 
\ion{C}{ii} lines are also present in the early epochs of both SNe. We find 
agreement with \cite{Brown2019} that the \ion{C}{ii} lines are present at early 
times in SN~2017erp, but disappear in our fits after $\sim -10$ days. We 
similarly find evidence for a weak \ion{C}{ii} line in the $-13.9$ day spectra 
of SN~2021fxy, which disappears before our next spectrum at $-5.9$ days. Both 
SNe show features at the expected location of the \ion{O}{i} $\lambda7773$ line, 
yet the contamination of this feature by telluric lines and a strong contribution 
of \ion{Mg}{ii} $\lambda\lambda 7896,7877$ doublet (likely overemphasized by our 
\texttt{SYNOW} fit) in SN~2021fxy makes fitting difficult. As a result, we can 
only definitively identify \ion{O}{i} in the $+5.1$ day spectrum. In order to fit 
the blue \ion{Mg}{ii} features in SNe~2021fxy and 2017erp, our fits require that 
the \ion{Mg}{ii} be located at high velocities and/or have high excitation 
temperatures, resulting in the high excitation lines at $\lambda\lambda7896,7877$ 
appearing abnormally strong at the location of the \ion{O}{i} lines. Examination 
of the NIR spectra of SN~2021fxy reveals \ion{Mg}{ii} lines that are weaker than 
those found in other NIR spectra of other SNe~Ia, likely caused by the MUV 
suppression preventing the UV photons from exciting the upper states of \ion{Mg}{ii}. 

At early times, the influence of IGEs on the spectra is restricted to HV 
and PV components of \ion{Fe}{ii} and \ion{Fe}{iii}. In SN~2017erp, \ion{Fe}{iii} 
is photospheric at all epochs except $-17.0$ days, while \ion{Fe}{ii} is 
consistently found as a high velocity feature. In SN~2021fxy however, the HV 
\ion{Fe}{iii} persists until at least the $-5.9$ day spectrum, while a weak 
photospheric component of \ion{Fe}{ii} begins appearing as early as $-5.9$ 
days. At later epochs, the influence of \ion{Ni}{ii} and \ion{Co}{ii} on the 
spectra become stronger, as the photosphere recedes into the Fe-rich inner 
regions of the ejecta. 

\clearpage

\begin{table*}
\caption{SN~2017erp \texttt{SYNOW} Parameters \label{tab:synow_17erp}}
\begin{tabular}{ccccccccc}
  \hline
  Ion & Parameter & $-17.0$ d & $-14.1$ d & $-8.5$ d & $-1.0$ d & 0.0 d & $+6.0$ d & $+11.4$ d \\
  \hline
  & $T_{bb}$ [K] & 10500 & 10500 & 9500 & 11000 & 11000 & 10000 & 9800 \\*
  & $v_{phot}$ [$10^3$ km s$^{-1}$] & 15.0 & 13.9 & 12.5 & 11.0 & 11.0 & 9.8 & 8.5 \\*
  & $v_{max}$ [$10^3$ km s$^{-1}$] & 45.0 & 45.0 & 35.0 & 30.0 & 25.0 & 25.0 & 25.0 \\
  \hline
  \multirow{4}{*}{C II}
  & $\tau$ & --- & 0.14 & --- & --- & --- & --- & --- \\*
  & $v_{min}$/$v_{max}$ & --- & 13.9/25.0 & --- & --- & --- & --- & --- \\*
  & $v_{e}$ & --- & 2.0 & --- & --- & --- & --- & --- \\*
 & $T_{exec}$ & --- & 20000 & --- & --- & --- & --- & ---  \\
 \hline
\end{tabular} \\
{\it Note:} The full table of \texttt{SYNOW} parameters for the SN~2017erp fits shown in 
\autoref{fig:synow_fits} are available online as supplemental material.
\end{table*}

\begin{table*}
\caption{SN~2021fxy \texttt{SYNOW} Parameters \label{tab:synow_21fxy}}
\begin{tabular}{cccccccc}
  \hline
  Ion & Parameter & $-13.9$ d & $-5.9$ d & $+1.3$ d & $+3.2$ d & $+5.1$ d & $+10.9$ d \\
  \hline
  & $T_{bb}$ [K] & 10500 & 13000 & 13000 & 12000 & 15000 & 9300\\
  & $v_{phot}$ [$10^3$ km s$^{-1}$] & 15.0 & 10.8 & 10.0 & 10.0 & 9.0 & 8.0\\
  & $v_{max}$ [$10^3$ km s$^{-1}$] & 40.0 & 30.0 & 25.0 & 25.0 & 25.0 & 25.0\\
  \hline
  \multirow{4}{*}{C II}
 & $\tau$ & 0.18 & --- & --- & --- & --- & ---\\
 & $v_{min}$/$v_{max}$ & 15.0/25.0 & --- &   --- & --- & --- & ---\\
 & $v_{e}$ & 2.0 & --- & --- & --- & --- & ---\\
 & $T_{exec}$ & 17000 & --- & --- & --- & --- & ---\\
  \hline 
\end{tabular} \\
{\it Note:} The full table of \texttt{SYNOW} parameters for the SN~2021fxy fits shown in 
\autoref{fig:synow_fits} are available online as supplemental material.
\end{table*}

\begin{figure*}
  \centering
  \includegraphics[width=\textwidth]{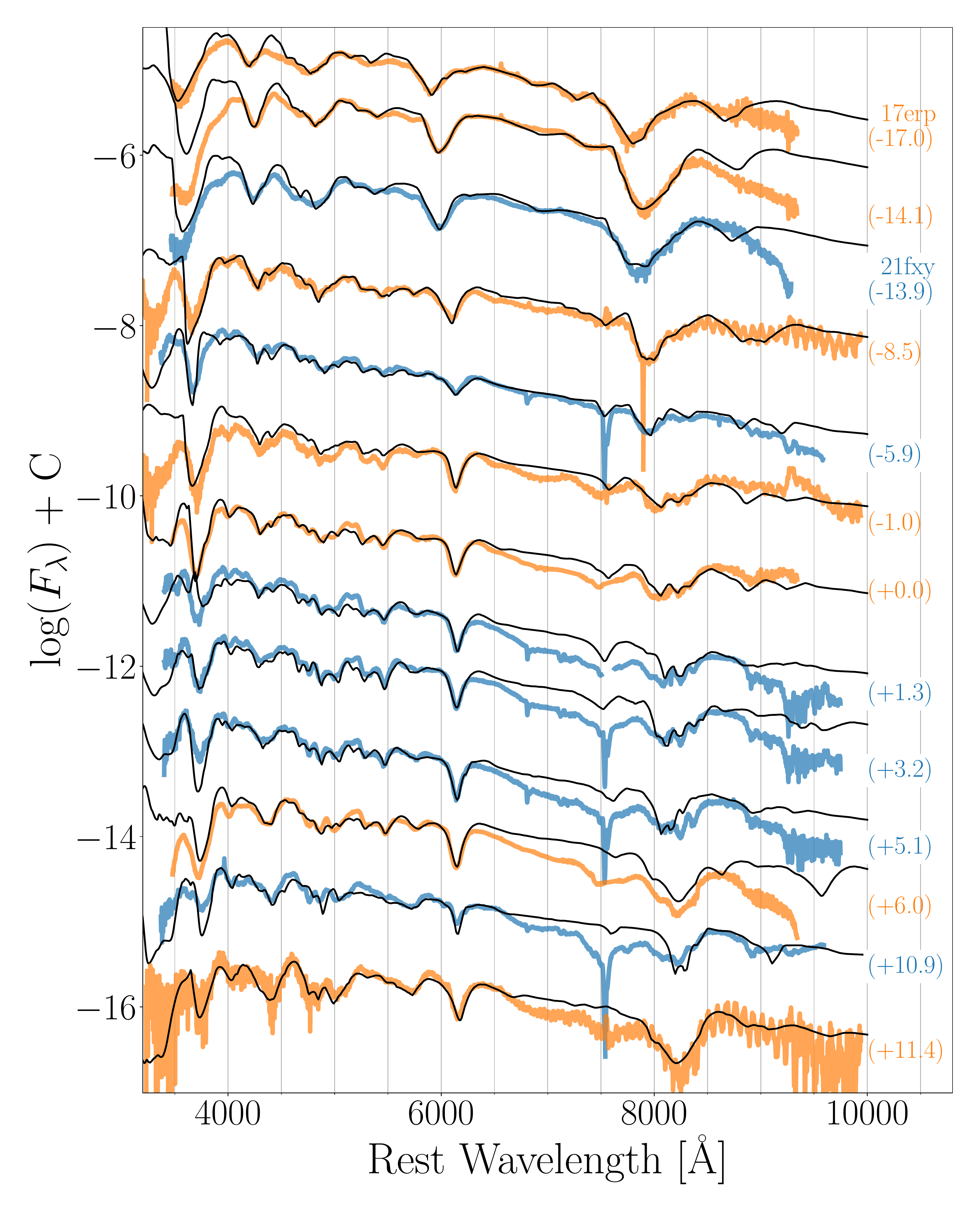}
  \caption{Optical sequences of SN~2017erp (orange) and SN~2021fxy (blue), with 
  \texttt{SYNOW} fits overlaid in black.}
  \label{fig:synow_fits}
\end{figure*}


\bsp	
\label{lastpage}
\end{document}


\begin{deluxetable}{@{\extracolsep{0pt}}ccccccccc}
  \centering
  \tabletypesize{\footnotesize}
  \tablewidth{0pt}
  \tablecaption{SN 2017erp \texttt{SYNOW} Parameters \label{tab:synow_17erp}}
  \tablehead{\colhead{Ion} & \colhead{Parameter} & \colhead{-17.0 d} & \colhead{-14.1 d} &
  \colhead{-8.5 d} & \colhead{-1.0 d} & \colhead{-0.0 d} & 
  \colhead{+6.0 d} & \colhead{+11.4 d}}
  \startdata
  & $T_{bb}$ [K] & 10500 & 10500 & 9500 & 11000 & 11000 & 10000 & 9800 \\
  & $v_{phot}$ [$10^3$ km s$^{-1}$] & 15.0 & 13.9 & 12.5 & 11.0 & 11.0 & 9.8 & 8.5 \\
  & $v_{max}$ [$10^3$ km s$^{-1}$] & 45.0 & 45.0 & 35.0 & 30.0 & 25.0 & 25.0 & 25.0 \\
  \hline
  \multirow{4}{*}{C II}
  & $\tau$ & \nodata & 0.14 & \nodata & \nodata & \nodata & \nodata & \nodata \\
  & $v_{min}$/$v_{max}$ & \nodata & 13.9/25.0 & \nodata & \nodata & \nodata & \nodata & \nodata \\
  & $v_{e}$ & \nodata & 2.0 & \nodata & \nodata & \nodata & \nodata & \nodata \\
  & $T_{exec}$ & \nodata & 20000 & \nodata & \nodata & \nodata & \nodata & \nodata  \\
  \hline
  \multirow{4}{*}{HV C II}
  & $\tau$ & 0.15 & \nodata & \nodata & \nodata & \nodata & \nodata & \nodata \\
  & $v_{min}$/$v_{max}$ & 17.5/25.0 & \nodata & \nodata & \nodata & \nodata & \nodata & \nodata \\
  & $v_{e}$ & 2.0 & \nodata & \nodata & \nodata & \nodata & \nodata & \nodata \\
  & $T_{exec}$ & 20000 & \nodata & \nodata & \nodata & \nodata & \nodata & \nodata  \\
  \hline
  \multirow{4}{*}{O I}
  & $\tau$ & \nodata & \nodata & \nodata & \nodata & \nodata & 0.20 & \nodata \\
  & $v_{min}$/$v_{max}$ & \nodata & \nodata & \nodata & \nodata & \nodata & 9.8/25.0 & \nodata  \\
  & $v_{e}$ & \nodata & \nodata & \nodata & \nodata & \nodata & 1.0 & \nodata  \\
  & $T_{exec}$ & \nodata & \nodata & \nodata & \nodata & \nodata & 20000 & \nodata \\
  \hline
  \multirow{4}{*}{HV O I}
  & $\tau$ & 0.30 & 0.15 & \nodata & \nodata & 0.20 & \nodata & \nodata \\
  & $v_{min}$/$v_{max}$ & 21.0/30.0 & 18.0/25.0 & \nodata & \nodata & 13.0/25.0 & \nodata & \nodata  \\
  & $v_{e}$ & 4.0 & 4.0 & \nodata & \nodata & 2.0 & \nodata & \nodata \\
  & $T_{exec}$ & 10000 & 10000 & \nodata & \nodata & 15000 & \nodata & \nodata \\
  \hline
  \multirow{4}{*}{Na I}
  & $\tau$ & \nodata & \nodata & \nodata & \nodata & \nodata & 0.4 & 0.8 \\
  & $v_{min}$/$v_{max}$ & \nodata & \nodata & \nodata & \nodata & \nodata & 9.8/25.0 & 8.5/25.0 \\
  & $v_{e}$ & \nodata & \nodata & \nodata & \nodata & \nodata & 1.0 & 5.0 \\
  & $T_{exec}$ & \nodata & \nodata & \nodata & \nodata & \nodata & 15000 & 15000 \\
  \hline
  \multirow{4}{*}{Mg II}
  & $\tau$ & 1.0 & 1.0 & \nodata  & \nodata & \nodata & 0.6 & 1.0 \\
  & $v_{min}$/$v_{max}$ & 15.0/25.0 & 13.9/25.0 & \nodata & \nodata & \nodata & 10.0/25.0 & 8.5/25.0 \\
  & $v_{e}$ & 3.0 & 3.0 & \nodata & \nodata & \nodata & 3.0 & 0.50 \\
  & $T_{exec}$ & 10000 & 10000 & \nodata & \nodata & \nodata & 15000 & 15000 \\
  \hline
  \multirow{4}{*}{HV Mg II}
  & $\tau$ & 1.4 & 2.5 & 1.0 & 1.1 & 1.5 & \nodata &  \nodata \\
  & $v_{min}$/$v_{max}$ & 20.0/35.0 & 16.0/30.0 & 14.0/25.0 & 13.0/25.0 & 13.0/25.0 & \nodata &  \nodata \\
  & $v_{e}$ & 5.0 & 5.0 & 5.0 & 2.0 & 2.0 & \nodata & \nodata \\
  & $T_{exec}$ & 5000 & 5000 & 8000 & 15000 & 20000 & \nodata & \nodata \\
  \hline
  \multirow{4}{*}{Si II}
  & $\tau$ & 0.60 & 1.2 & 8.0 & 4.0 & 5.0 & 5.0 & 4.0 \\
  & $v_{min}$/$v_{max}$ & 15.0/25.0 & 13.9/25.0 & 12.5/25.0 & 11.0/15.0 & 11.0/15.0 & 9.8/25.0 & 8.5/25.0 \\
  & $v_{e}$ & 5.0 & 7.0 & 2.0 & 2.0 & 2.0 & 2.0 & 2.0 \\
  & $T_{exec}$ & 10000 & 10000 & 10000 & 10000 & 11000 & 11000 & 10000 \\
  \hline
  \multirow{4}{*}{HV Si II}
  & $\tau$ & 1.0 & 1.4 & 1.4 & 0.70 & 1.5 & \nodata & \nodata \\
  & $v_{min}$/$v_{max}$  & 23.0/35.0 & 20.0/28.0 & 14.0/25.0 & 14.0/25.0 & 13.0/25.0 & \nodata & \nodata \\
  & $v_{e}$ & 5.0 & 5.0 & 4.0 & 1.0 & 1.0 & \nodata & \nodata \\
  & $T_{exec}$ & 10000 & 10000 & 20000 & 10000 & 10000 & \nodata & \nodata  \\
  \hline
  \multirow{4}{*}{Si III}
  & $\tau$ & \nodata & \nodata & 4.0 & 1.9 & 2.0 & 1.0 & 5.0 \\
  & $v_{min}$/$v_{max}$ & \nodata & \nodata & 13.0/25.0 & 11.0/25.0 & 11.0/25.0 & 9.8/25.0 & 8.5/25.0 \\
  & $v_{e}$ & \nodata & \nodata & 0.50 & 1.0 & 1.0 & 2.0 & 0.80 \\
  & $T_{exec}$ & \nodata & \nodata & 25000 & 5000 & 5000 & 5000 & 10000 \\
  \hline
  \multirow{4}{*}{HV Si III}
  & $\tau$ & 0.5 & 0.5 & \nodata & \nodata & \nodata & \nodata & \nodata \\
  & $v_{min}$/$v_{max}$& 19.0/25.0 & 16.0/25.0 & \nodata & \nodata & \nodata & \nodata & \nodata \\
  & $v_{e}$ & 3.0 & 3.0 & \nodata & \nodata & \nodata & \nodata & \nodata  \\
  & $T_{exec}$ & 10000 & 10000 & \nodata & \nodata & \nodata & \nodata & \nodata \\
  \hline
  \multirow{3}{*}{S II}
  & $\tau$ & \nodata & 1.1 & 1.8 & 1.7 & 1.3 & 2.4 & 0.30 \\
  & $v_{min}$/$v_{max}$ & \nodata & 13.9/25.0 & 12.5/25.0 & 11.0/25.0 & 11.0/25.0 & 9.8/25.0 & 8.5/25.0 \\
  & $v_{e}$ & \nodata & 2.0 & 0.80 & 1.0 & 2.0 & 1.0 & 3.0 \\
  & $T_{exec}$ & \nodata & 15000 & 10000 & 10000 & 15000 & 16000 & 20000 \\
  \hline
  \multirow{4}{*}{HV S II}
  & $\tau$ & 0.90 & \nodata & \nodata & \nodata & \nodata & \nodata & \nodata \\
  & $v_{min}$/$v_{max}$ & 18.0/25.0 & \nodata & \nodata & \nodata & \nodata & \nodata & \nodata \\
  & $v_{e}$ & 2.0 & \nodata & \nodata & \nodata & \nodata & \nodata & \nodata \\
  & $T_{exec}$ & 25000 & \nodata & \nodata & \nodata & \nodata & \nodata & \nodata \\
  \hline
  \multirow{4}{*}{Ca II}
  & $\tau$ & 18 & \nodata & 15 & 10(11) & 25 & 70 & 100 \\
  & $v_{min}$/$v_{max}$ & 16.0/25.0 & \nodata & 12.5/25.0 & 11.0/15.0 (12.0/20.0) & 12.0/15.0 & 9.8/25.0 & 8.5/23.0 \\
  & $v_{e}$ & 7.0 & \nodata & 3.0 & 3.0(2.0) & 2.0 & 3.0 & 4.0 \\
  & $T_{exec}$ & 8000 & \nodata & 10000 & 11500 & 11000 & 20000 & 10000 \\
  \hline
  \multirow{4}{*}{HV Ca II}
  & $\tau$ & 30 & 120 & 12 & 7.0 & 8.0 & 30 & \nodata \\
  & $v_{min}$/$v_{max}$ & 24.0/30.0 & 20.0/30.0 & 19.0/25.0 & 18.0/30.0 & 16.5/25.0 & 12.0/25.0 & \nodata \\
  & $v_{e}$ & 7.0 & 8.0 & 7.0 & 6.0 & 10.0 & 3.0 & \nodata \\
  & $T_{exec}$ & 8000 & 10000 & 10000 & 11500 & 10000 & 15000 & \nodata \\
  \hline
  & $\tau$ & 35 & 50 & 50 & \nodata & \nodata & \nodata & \nodata \\
  & $v_{min}$/$v_{max}$ & 29.0/45.0 & 23.0/36.5 & 20.0/30.0 & \nodata & \nodata & \nodata & \nodata \\
  & $v_{e}$ & 7.0 & 7.0 & 5.0 & \nodata & \nodata & \nodata & \nodata \\
  & $T_{exec}$ & 8000 & 10000 & 6000 & \nodata & \nodata & \nodata & \nodata \\
  \hline
  \multirow{4}{*}{Fe II}
  & $\tau$ & \nodata & \nodata & 1.0 & 0.30 & \nodata & \nodata & \nodata\\
  & $v_{min}$/$v_{max}$ & \nodata & \nodata & 13.5/25.0 & 11.0/15.0 & \nodata & \nodata & \nodata\\
  & $v_{e}$ & \nodata & \nodata & 1.0 & 5.0 & \nodata & \nodata & \nodata\\
  & $T_{exec}$ & \nodata & \nodata & 10000 & 11000 & \nodata & \nodata & \nodata\\
  \hline
  \multirow{4}{*}{HV Fe II}
  & $\tau$ & 0.30(0.60) & 0.60 & \nodata & \nodata & 0.40 & 0.80 & 3.0 \\
  & $v_{min}$/$v_{max}$ & 19.0/25.0(25.0/35.0) & 22.0/30.0 & \nodata & \nodata & 17.0/25.0 & 12.0/25.0 & 11.0/25.0 \\
  & $v_{e}$ & 2.0(10.0) & 10.0 & \nodata & \nodata & 1.0 & 2.0 & 2.0 \\
  & $T_{exec}$ & 10000(8000) & 8000 & \nodata & \nodata & 12000 & 10000 & 10000 \\
  \hline
  \multirow{4}{*}{Fe III}
  & $\tau$ & \nodata & \nodata & 0.40 & \nodata & 0.10 & 1.0 & 2.0\\
  & $v_{min}$/$v_{max}$ & \nodata & \nodata & 13.0/25.0 & \nodata & 12.0/25.0 & 9.8/25.0 & 8.5/25.0 \\
  & $v_{e}$ & \nodata & \nodata & 2.0 & \nodata & 1.0 & 1.0 & 1.0\\
  & $T_{exec}$ & \nodata & \nodata & 15000 & \nodata & 10000 & 15000 & 8000\\
  \hline
  \multirow{4}{*}{HV Fe III}
  & $\tau$ & 0.90 & 0.60 & \nodata & 0.23 & \nodata & \nodata & \nodata\\
  & $v_{min}$/$v_{max}$ & 20.0/25.0 & 16.0/25.0 & \nodata & 13.0/25.0 & \nodata & \nodata & \nodata\\
  & $v_{e}$ & 1.0 & 1.5 & \nodata & 1.0 & \nodata & \nodata & \nodata\\
  & $T_{exec}$ & 8000 & 8000 & \nodata & 8000 & \nodata & \nodata & \nodata\\
  \hline
  \multirow{4}{*}{Co II}
  & $\tau$ & \nodata & \nodata & \nodata & \nodata & 1.5 & 1.0 & 3.0\\
  & $v_{min}$/$v_{max}$ & \nodata & \nodata & \nodata & \nodata & 12.0/25.0 & 10.0/25.0 & 8.5/25.0\\
  & $v_{e}$ & \nodata & \nodata & \nodata & \nodata & 1.0 & 1.0 & 4.0\\
  & $T_{exec}$ & \nodata & \nodata & \nodata & \nodata & 5000 & 1000 & 7000\\
  \enddata
  \tablecomments{Values in parenthesis indicate a second detached component to the feature.}
\end{deluxetable}

\begin{deluxetable}{@{\extracolsep{0pt}}cccccccc}
  \centering
  \tablewidth{0pt}
  \tablecaption{SN 2021fxy \texttt{SYNOW} Parameters \label{tab:synow_21fxy}}
  \tablehead{\colhead{Ion} & \colhead{Parameter} & \colhead{-14.0 d} & \colhead{-6.0 d} &
  \colhead{+1.3 d} & \colhead{+3.2 d} & \colhead{+5.1 d} & \colhead{+11.0 d}}
  \startdata
  & $T_{bb}$ [K] & 10500 & 13000 & 13000 & 12000 & 15000 & 9300\\
  & $v_{phot}$ [$10^3$ km s$^{-1}$] & 15.0 & 10.8 & 10.0 & 10.0 & 9.0 & 8.0\\
  & $v_{max}$ [$10^3$ km s$^{-1}$] & 40.0 & 30.0 & 25.0 & 25.0 & 25.0 & 25.0\\
  \hline 
  \multirow{4}{*}{C II}
  & $\tau$ & 0.18 & \nodata & \nodata & \nodata & \nodata & \nodata\\
  & $v_{min}$/$v_{max}$ & 15.0/25.0 & \nodata &   \nodata & \nodata & \nodata & \nodata\\
  & $v_{e}$ & 2.0 & \nodata & \nodata & \nodata & \nodata & \nodata\\
  & $T_{exec}$ & 17000 & \nodata & \nodata & \nodata & \nodata & \nodata\\
  \hline 
  \multirow{4}{*}{HV C II}
  & $\tau$  & \nodata &  \nodata & \nodata & \nodata & 0.017 & \nodata\\
  & $v_{min}$/$v_{max}$ & \nodata & \nodata & \nodata &\nodata & 17.0/25.0 & \nodata \\
  & $v_{e}$ & \nodata & \nodata & \nodata & \nodata & 0.50 & \nodata\\
  & $T_{exec}$ & \nodata  & \nodata & \nodata & \nodata & 10000 & \nodata  \\
  \hline
  \multirow{4}{*}{O I}
  & $\tau$ & \nodata & \nodata & \nodata & \nodata & 0.30 & \nodata\\
  & $v_{min}$/$v_{max}$ & \nodata & \nodata & \nodata & \nodata & 9.0/25.0 & \nodata\\
  & $v_{e}$ & \nodata & \nodata & \nodata & \nodata & 2.0 & \nodata\\
  & $T_{exec}$ & \nodata & \nodata & \nodata & \nodata & 10000 & \nodata\\
  \hline
  \multirow{4}{*}{Na I}
  & $\tau$ & \nodata & 1.0 & 0.50 & 0.50 & 0.20 & 0.30\\
  & $v_{min}$/$v_{max}$ & \nodata & 10.8/25.0 & 10.0/25.0 & 10.0/25.0 & 10.0/25.0 & 9.0/25.0\\
  & $v_{e}$ & \nodata & 0.6 & 2.0 & 2.0 & 3.0 & 6.0\\
  & $T_{exec}$ & \nodata & 8000 & 10000 & 10000 & 10000 & 5000\\
  \hline
  \multirow{4}{*}{Mg II}
  & $\tau$ & \nodata & \nodata & \nodata & \nodata & 0.60 & \nodata\\
  & $v_{min}$/$v_{max}$ & \nodata & \nodata & \nodata & \nodata & 11.0/25.0 & \nodata\\
  & $v_{e}$ & \nodata & \nodata & \nodata & \nodata & 2.0 & \nodata\\
  & $T_{exec}$ & \nodata & \nodata & \nodata & \nodata & 10000 & \nodata\\
  \hline
  \multirow{4}{*}{HV Mg II}
  & $\tau$ & 1.2 & 0.55 & 0.5 & 0.01 & 0.10 & \nodata\\
  & $v_{min}$/$v_{max}$ & 18.0/30.0 & 15.0/30.0 & 12.0/25.0 & 13.0/25.0 & 18.0/25.0 & \nodata\\
  & $v_{e}$ & 4.0 & 5.0 & 3.0 & 4.0 & 2.0 & \nodata\\
  & $T_{exec}$ & 5000 & 15000 & 15000 & 15000 & 20000 & \nodata \\
  \hline
  \multirow{4}{*}{Si II}
  & $\tau$ & 0.80 & 0.80 & 2.85 & 2.3 & 3.0 & 0.25\\
  & $v_{min}$/$v_{max}$ & 15.0/25.0 & 10.8/30.0 & 10.0/25.0 & 10.0/25.0 & 10.0/25.0 & 8.0/25.0\\
  & $v_{e}$ & 4.0 & 2.5 & 2.0 & 2.0 & 1.7 & 2.0\\
  & $T_{exec}$ & 10000 & 10000 & 10000 & 10000 & 10000 & 10000\\
  \hline
  \multirow{4}{*}{HV Si II}
  & $\tau$ & 1.6 & 0.15 & 0.10 & 0.05 & \nodata & 2.0\\
  & $v_{min}$/$v_{max}$ & 19.0/30.0 & 21.0/30.0 & 20.0/25.0 & 22.0/25.0 & \nodata & 10.5/25.0\\
  & $v_{e}$ & 5.0 & 0.5 & 3.0 & 3.0 & \nodata & 1.2\\
  & $T_{exec}$ & 5000 & 10000 & 8000 & 8000 & \nodata & 20000\\
  \hline
  \multirow{4}{*}{Si III}
  & $\tau$ & 2.3 & 1.5 & 0.80 & 0.50 & 0.10 & \nodata\\
  & $v_{min}$/$v_{max}$ & 15.0/30.0 & 10.8/30.0 & 10.0/25.0 & 10.0/25.0 & 10.0/25.0 & \nodata\\
  & $v_{e}$ & 3.0 & 1.5 & 1.0 & 1.0 & 1.0 & \nodata\\
  & $T_{exec}$ & 10000 & 18000 & 5000 & 12000 & 12000 & \nodata\\
  \hline
  \multirow{4}{*}{HV Si III}
  & $\tau$ & \nodata & \nodata & \nodata & \nodata & \nodata & 0.4 \\
  & $v_{min}$/$v_{max}$ & \nodata & \nodata & \nodata & \nodata & \nodata & 10.0/25.0 \\
  & $v_{e}$ & \nodata & \nodata & \nodata & \nodata & \nodata & 2.0 \\
  & $T_{exec}$ & \nodata & \nodata & \nodata & \nodata & \nodata & 8000 \\
  \hline
  \multirow{4}{*}{S II}
  & $\tau$ & 1.2 & 0.80 & 1.6 & 4.5 & 2.2 & 0.05\\
  & $v_{min}$/$v_{max}$ & 16.0/30.0 & 10.8/30.0 & 10.0/25.0 & 10.0/25.0 & 10.0/25.0 & 9.0/25.0\\
  & $v_{e}$ & 2.0 & 1.5 & 2.0 & 1.0 & 1.0 & 0.5\\
  & $T_{exec}$ & 10000 & 16000 & 28000 & 30000 & 25000 & 5000\\
  \hline
  \multirow{4}{*}{Ca II}
  & $\tau$ & 15.0 & 1.0 & 17.0 & 21.0 & 15.0 & 80.0\\
  & $v_{min}$/$v_{max}$ & 15.0/25.0 & 10.8/30.0 & 10.0/25.0 & 10.0/25.0 & 10.0/25.0 & 10.0/25.0 \\
  & $v_{e}$ & 6.0 & 2.0 & 2.0 & 2.0 & 1.0 & 1.0\\
  & $T_{exec}$ & 10000 & 15000 & 15000 & 15000 & 5000 & 10000\\
  \hline
  \multirow{4}{*}{HV Ca II}
  & $\tau$ & 23.0 & 5.0 & 37.0 & 55.0 & 16.0 & 30.0\\
  & $v_{min}$/$v_{max}$ & 25.0/40.0 & 22.0/30.0 & 20.0/25.0 & 18.5/25.0 & 15.0/25.0 & 13.0/25.0\\
  & $v_{e}$ & 10.0 & 5.0 & 2.0 & 1.0 & 2.0 & 4.0\\
  & $T_{exec}$ & 12000 & 15000 & 12000 & 10000 & 10000 & 10000\\
  \hline
  & $\tau$ & 20.0 & \nodata & \nodata & \nodata & 22.0 & 10.0\\
  & $v_{min}$/$v_{max}$ & 23.0/30.0 & \nodata & \nodata & \nodata & 18.0/25.0 & 15.0/25.0\\
  & $v_{e}$ & 10.0 & \nodata & \nodata & \nodata & 2.0 & 2.0\\
  & $T_{exec}$ & 10000 & \nodata & \nodata & \nodata & 10000 & 8000\\
  \hline
  \multirow{4}{*}{Fe II}
  & $\tau$ & \nodata & 0.10 & \nodata & 0.30 & 0.70 & 1.0\\
  & $v_{min}$/$v_{max}$ & \nodata & 10.8/30.0 & \nodata & 10.0/25.0 & 10.0/25.0 & 8.0/25.0\\
  & $v_{e}$ & \nodata & 2.0 & \nodata & 2.0 & 2.8 & 2.0\\
  & $T_{exec}$ & \nodata & 10000 & \nodata & 10000 & 10000 & 10000\\
  \hline
  \multirow{4}{*}{HV Fe II}
  & $\tau$ & 1.0(1.5) & 0.50 & 0.60 & 0.80 & 0.50 & \nodata\\
  & $v_{min}$/$v_{max}$ & 22.0/30.0(23.5/35.0) & 16.5/30.0 & 17.0/25.0 & 19.0/25.0 & 18.0/25.0 & \nodata\\
  & $v_{e}$ & 4.0(3.0) & 1.0 & 2.0 & 2.0 & 1.5 & \nodata\\
  & $T_{exec}$ & 10000(9000) & 10000 & 11000 & 8000 & 10000 & \nodata\\
  \hline
  \multirow{4}{*}{Fe III}
  & $\tau$ & 1.5 & 0.30 & 0.10 & 0.10 & 0.50 & \nodata\\
  & $v_{min}$/$v_{max}$ & 16.0/30.0 & 10.8/30.0 & 10.0/25.0 & 10.0/25.0 & 9.0/25.0 & \nodata\\
  & $v_{e}$ & 2.0 & 4.0 & 2.0 & 2.0 & 2.0 & \nodata\\
  & $T_{exec}$ & 10000 & 15000 & 10000 & 5000 & 15000 & \nodata\\
  \hline
  \multirow{4}{*}{HV Fe III}
  & $\tau$ & \nodata & 0.20 & \nodata & \nodata & \nodata & \nodata\\
  & $v_{min}$/$v_{max}$ & \nodata & 15.0/30.0 & \nodata & \nodata & \nodata & \nodata\\
  & $v_{e}$ & \nodata & 4.0 & \nodata & \nodata & \nodata & \nodata\\
  & $T_{exec}$ & \nodata & 15000 & \nodata & \nodata & \nodata & \nodata\\
  \hline
  \multirow{4}{*}{Co II}
  & $\tau$ & \nodata & \nodata & 0.80 & 2.0 & 1.0 & 2.0\\
  & $v_{min}$/$v_{max}$ & \nodata & \nodata & 10.0/25.0 & 10.0/25.0 & 9.0/25.0 & 8.0/25.0\\
  & $v_{e}$ & \nodata & \nodata & 1.0 & 2.0 & 2.0 & 2.0\\
  & $T_{exec}$ & \nodata & \nodata & 7000 & 7000 & 7000 & 7000\\
  \hline
  \multirow{2}{*}{Ni II}
  & $\tau$ & \nodata & 0.50 & 1.0 & 0.50 & 1.0 & \nodata\\
  & $v_{min}$/$v_{max}$ & \nodata & 10.8/30.0 & 10.0/25.0 & 10.0/25.0 & 9.0/25.0 & \nodata\\
  & $v_{e}$ & \nodata & 3.0 & 3.0 & 3.0 & 2.0 & \nodata\\
  & $T_{exec}$ & \nodata & 4500 & 4500 & 4500 & 7000 & \nodata\\
  \enddata
  \tablecomments{Values in parenthesis indicate a second detached component to the feature.}
\end{deluxetable}